\documentclass[12pt,a4paper]{article}

\usepackage{amsmath,amsfonts,latexsym,amssymb}
\usepackage{verbatim,amsthm,curves,graphics}
\usepackage{mathrsfs}

\def\ben{\begin{equation}}
\def\een{\end{equation}}
\def\bena{\begin{eqnarray}}
\def\eena{\end{eqnarray}}

\def\f(#1/#2){\frac{#1}{#2}} 
\def\Frac(#1/#2){\left(\frac{#1}{#2}\right)} 
\def\chris(#1-#2-#3){{\mit \Gamma}^{#1}{}_{{#2}{#3}} }
\def\tilchris(#1-#2-#3){\tilde{{\mit \Gamma}}^{#1}{}_{{#2}{#3}}}
\def\hatchris(#1-#2-#3){\hat{{\mit \Gamma}}^{#1}{}_{{#2}{#3}}}

\newcommand{\non}{\nonumber}
\theoremstyle{definition}

\newcommand{\I}{{\mathscr I}}
\renewcommand{\S}{{\mathscr S}}

\newcommand{\tnabla}{{\tilde \nabla}}
\newcommand{\tg}{{\tilde g}}

\renewcommand{\d}{{\rm d}}

\begin{document}

\title{Asymptotic Flatness and Bondi Energy in Higher Dimensional Gravity} 
\author{Stefan Hollands\thanks{Electronic mail: \tt stefan@bert.uchicago.edu}\\
                    \it{Enrico Fermi Institute, Department of Physics,}\\
                    \it{University of Chicago, 5640 Ellis Ave.,}\\
                    \it{Chicago IL 60637, USA} \\
\\
       Akihiro Ishibashi\thanks{Electronic mail: \tt
                                A.Ishibashi@damtp.cam.ac.uk}\\
                    \it{Department of Applied Mathematics and
                        Theoretical Physics,} \\ 
                    \it{Centre for Mathematical Sciences,} \\
                    \it{University of Cambridge, Wilberforce Road,}\\ 
                    \it{Cambridge CB3 0WA, UK} \\  
        }
\date{\today}

\maketitle 

\begin{abstract}
We give a general geometric definition of asymptotic flatness at null 
infinity in $d$-dimensional general relativity ($d$ even) within the 
framework of conformal infinity. Our definition is arrived 
at via an analysis of linear perturbations near null infinity and 
shown to be stable under such perturbations. 
The detailed fall off properties of the perturbations, as 
well as the gauge conditions that need to be imposed to make the 
perturbations regular at infinity, are qualitatively different in 
higher dimensions; in particular, 
the decay rate of a radiating solution at null infinity differs from 
that of a static solution in higher dimensions. The definition of 
asymptotic flatness in higher dimensions consequently also differs
qualitatively from that in $d=4$.

We then derive an expression for the generator 
conjugate to an asymptotic time translation symmetry 
for asymptotically flat spacetimes in $d$-dimensional general
relativity ($d$ even) within the Hamiltonian framework, 
making use especially of a formalism 
developed by Wald and Zoupas. This generator is given 
by an integral over a cross section at null infinity of a certain 
local expression and is taken to be the definition of the Bondi 
energy in $d$ dimensions. Our definition yields 
a manifestly positive flux of radiated energy.  

Our definitions and constructions fail in odd spacetime dimensions, 
essentially because the regularity properties of the metric at null infinity
seem to be insufficient in that case. We also find that there is no
direct analog of the well-known infinite set of angle dependent translational
symmetries in more than 4 dimensions. 

\end{abstract}

\section{Introduction}
\label{sect:1}

Theories attempting to unify the forces often require 
a higher dimensional spacetime,  
and accordingly have different properties than 
$4$-dimensional theories. Still important and fundamental 
even in higher dimensional theories are the notion of 
an isolated system and associated conserved quantities, 
such as the total energy of the system. 

In 4-dimensional general relativity, there exist two satisfactory 
notions of the total energy of a spacetime representing an isolated
system, namely the ``Arnowitt-Deser-Misner'' (ADM) energy~\cite{adm} 
and the ``Bondi'' energy~\cite{bms,t,s}. 
The ADM energy represents the energy of the system 
``once and for all'' and is mathematically given by an integral of 
a quantity associated with the gravitational field, over a sphere 
at spatial infinity. The Bondi energy measures the total energy of 
the system ``at an instant of time'', and is mathematically given 
by an integral over a spacelike cross section at null infinity. 
Thus, while the ADM energy is just a number, the Bondi energy 
is in general a function of time in the sense that it depends on 
the chosen cross section at null infinity. The difference between 
the Bondi energies at two different times represents the flux 
of gravitational radiation through the portion of null infinity 
bounded by the corresponding two cross sections. 

While the expression for the ADM energy of a spacetime is readily 
generalized to an arbitrary number of spacetime dimensions, 
this is not so for the Bondi energy. To our knowledge, 
no expression for the Bondi energy or other quantities 
associated with the group of asymptotic symmetries has been 
given in the literature for more 
than $4$ dimensions\footnote{A generalization of 
the Bondi-Metzner-Sachs group to higher dimensions has been 
discussed in the context of supergravity~\cite{ags}. The definition of 
asymptotic flatness given in that paper differs from the definition given 
in the present paper.}, 
let alone a systematic derivation. This is maybe not so surprising since 
constructions associated with null infinity tend to be 
more complicated in nature than corresponding constructions 
at spatial infinity, which may e.g. be appreciated from the fact 
that a completely satisfactory definition of quantities associated 
with asymptotic symmetries in 4 dimensions was not given as late 
as the early 80's~\cite{ds}. The purpose of the present article 
is to derive an expression for the Bondi energy and momentum (Bondi energy, 
for short) in spacetimes of arbitrary (even) dimension. 

The basic issue that needs to be settled in order to even get started 
on a definition of Bondi energy in higher dimensions is to specify 
what exactly one means by the statement 
that a spacetime represents an ``isolated system''. 
Roughly speaking, an isolated system is a spacetime that looks like 
Minkowski space\footnote{Other, less restrictive notions of an isolated 
system may also be considered, for example systems that look like a 
Kaluza-Klein space far out in the ``non-compact directions''. However, the
analysis of such metrics and of the associated conserved quantities would be 
substantially different from the ones studied in the present paper.
} ``far away''. Of course, one has to 
explain exactly what one means by ``far away'', and one must   
determine the precise asymptotics that should  
reasonably be imposed on the gravitational field at null infinity. 
What are the asymptics for the gravitational field in $d$ dimensions?
For the sake of definiteness, 
suppose one would attempt to define an isolated system to be 
a spacetime whose metric has the form of the 
Schwarzschild metric (in suitable coordinates), plus higher order 
terms in $1/r$ as one goes off to infinity 
along a null direction. In other words, suppose one were to define 
asymptotic flatness at null infinity in the same way which works 
at spatial infinity. Then one would effectively eliminate from 
consideration all spacetimes that contain gravitational radiation 
through null infinity, which are of course precisely the spacetimes 
that one wants to describe in the first place. On the other hand, 
if one imposes drop off conditions that are to weak, 
then it will in general be impossible define a notion of radiated energy of 
such a spacetime in a meaningful way. Thus, the task is to find 
a definition of asymptotic flatness that is both general 
enough so as to allow sufficiently many physically interesting 
radiating spacetimes, and stringent enough so as to allow 
one to derive meaningful expressions for the energy, as well 
as possibly other quantities associated with asymptotic symmetries. 

The original definition of asymptotic flatness in 4 dimensions 
proposed by Bondi and collaborators~\cite{bms,s} was formulated 
in terms of detailed conditions on the metric components 
in a preferred coordinate frame and was arrived at via a study of 
gravitational waves near infinity. Their definition was later 
elegantly recast into the language of ``conformal infinity'' 
following the work of Penrose~\cite{p,tw,g1,g,as}. 
In this language, a spacetime is said to be 
asymptotically flat, if it can be conformally embedded into 
a smooth ``unphysical'' spacetime via a conformal 
transformation with conformal factor $\Omega$, so that the points at 
infinity are at the ``finite'' location $\Omega = 0$ in the unphysical 
spacetime, and so that the gradient of the conformal factor
$\Omega$ is null there. 
The main arguments that have been advanced in favor of 
this definition are that it covers the known exact solutions 
of Einstein's equation that one intuitively thinks of as representing 
isolated systems, and that the definition can be proven 
to be stable under linear perturbations~\cite{gx}, in the sense 
that any compactly generated solution to the linearized 
equations of motion around an asymptotically flat solution satisfies 
the linearized version of the above definition of asymptotic 
flatness in a suitable gauge\footnote{By contrast, if one were to adopt the 
e.g. same definition of asymptotic flatness at null infinity 
as one has at spatial infinity, then such a definition 
would not be stable under linear perturbations.}. 

Thus, the first task of 
our paper is to obtain an appropriate generalization 
of asymptotic flatness to higher dimensions. 
This definition will be motivated as in 4 dimensions by a
detailed analysis of the decay properties of linear perturbations 
at null infinity. We show that perturbations typically drop off 
as $1/r^{(d-2)/2}$ as one approaches null infinity, which, as 
we note, differs from the drop off rate of the Schwarzschild metric, 
$1/r^{d-3}$, when the spacetime dimensions is greater than 4. 
The appearance of half odd integer
powers of $1/r$ in the tail of the metric at null infinity in 
odd $d$ implies in particular that the unphysical metric will not 
be smooth at null infinity. It turns out that, for this reason, 
a geometrical definition of null infinity
as given above for $d=4$ does not appear to be possible in odd spacetime 
dimensions. We will therefore restrict our attention to spacetimes of even 
dimension in this paper. 
We will also see that  the detailed behavior of the perturbations 
near null infinity differs qualitatively from that in 4 dimensions 
in that the trace of the metric perturbation drops off one power 
in $1/r$ faster than the perturbation itself\footnote{As we will show, 
this phenomenon is closely 
related to the fact that the transverse traceless gauge is regular at  
null infinity in $d>4$, whereas this is not the case in $d=4$, where the 
so-called ``Geroch-Xanthopoulos'' gauge has this property.}. 
Consequently, our definition of asymptotic
flatness in $d > 4$ dimensions also differs qualitatively 
from that in 4 dimensions in that it involves, for example, 
additional conditions on the metric volume element as one 
approaches null infinity. 

The mathematical expression for the ADM energy, including 
the correct normalization, can be derived (in arbitrary dimension) 
in a simple and straightforward manner within the Hamiltonian 
framework of general relativity 
where it is seen to represent the ``charge'' conjugate to 
an infinitesimal asymptotic time translation at spatial 
infinity~\cite{rt}. It was shown by Wald and Zoupas~\cite{wz} 
(based on earlier work by Ashtekar and Streubel~\cite{as}, 
see also~\cite{abr}) 
that an expression for the Bondi energy in 4 dimensional 
general relativity can also be arrived at within a 
Hamiltonian framework as the quantity conjugate to an asymptotic time 
translation at null infinity, although the situation is certainly 
considerably more complicated compared to spatial infinity. 
This expression was shown to be unique under some natural assumptions 
and agrees with the previously known one~\cite{g, ds}. 
The formalism of~\cite{wz} is in fact capable of dealing 
with arbitrary diffeomorphism covariant theories of gravity 
(in arbitrary dimension) in the presence of boundary conditions 
at null-surfaces. We employ it here to establish 
the existence and uniqueness of a generator conjugate to an asymptotic
time-translation in $d$-dimensional vacuum general relativity ($d$ an even number) 
within the context of our asymptotic flatness condition, and we take this 
generator as the definition of the Bondi energy in $d$ dimensions. 
The algorithm by Wald and Zoupas specifies this generator 
only indirectly via its variation under a suitable variation 
of the spacetime metric, so further work is required to actually 
find a local expression for this quantity. Such an expression 
is provided in eq.~\eqref{bondimass}. As in 4 dimensions, our 
definition of the Bondi energy has the property that it yields 
a manifestly positive flux of 
energy given by the square of a suitably defined ``news tensor''. 

We emphasize that the issue of existence of a generator conjugate to an  
asymptotic time translation symmetry (or a more general other 
asymptotic symmetry) is by no means automatic but rather depends 
crucially on the nature of the boundary conditions. As a rule, boundary 
conditions that are ``too weak'' will prohibit the existence 
of a generator. In our case, the boundary conditions are independently 
determined by a perturbation analysis, and therefore not 
put in ``by hand'' in order to guarantee the existence of 
a generator. That we find existence of a generator is therefore 
a consistency check. 

The case $d=4$ seems 
to be ``exceptional'' with regard to many of our constructions 
from the point of view of general $d$. For example, 
the definition of the news tensor differs in dimensions 
greater than 4. Also, while the unphysical Weyl tensor can be 
proven to vanish one order faster than the unphysical Ricci tensor 
in $d=4$, this does not appear to be the case in higher dimensions. 
On the other hand, the unphysical Ricci scalar vanishes one order
faster than the unphysical Ricci tensor itself in $d>4$, while 
both quantities generically have the same drop off behavior in $d=4$.
In $d=4$, it is well known that the asymptotic symmetries form
an infinite dimensional group which comprises, besides the 
transformations corresponding to the usual symmtries of Minkowski 
spacetime, an additional infinite set of (mutually commuting) ``angle
dependent translations'', sometimes called 
``supertranslations''~\footnote{There is no relation with 
supersymmetry.}. We find that there is no direct analog  
of the supertranslations in $d>4$. Another curious feature, which is
of a more technical nature but nevertheless seems to underly 
many of the differences between $d=4$ and higher dimenions is that, 
while linear perturbations can be proven to 
be regular at null infinity in the transverse traceless gauge 
in $d>4$, this is not so in $d=4$, where 
one has to choose the (somewhat complicated) 
Geroch-Xanthopoulos gauge~\cite{gx}.

The contents and main results of this paper 
may now be summarized. In section~\ref{sect:2} we present our 
definition of asymptotic flatness in arbitrary dimension 
and define the notion of an asymptotic symmetry. 
In section~\ref{sect:3} we show that our definition of asymptotic flatness 
is stable under linear perturbations, provided that the 
perturbations are in a suitable gauge. In section~\ref{sect:4} we 
recall the general formalism of~\cite{wz} for defining generators 
associated with asymptotic symmetries, and in section~\ref{sect:5} we derive 
our formula for the Bondi energy.
We also verify that our expression for the Bondi energy agrees with 
the expected one (i.e., the ADM energy) in the $d$ dimensional 
analogue of the Schwarzschild metric. 
We draw our conclusions in section~\ref{sect:6}. 
For simplicity, we restrict attention to the case of vacuum general 
relativity throughout this article. Some remarks concerning 
the incorporation of matter fields are also given in section~\ref{sect:6}.

Our conventions are the same as in~\cite{w}: The signature 
of the metric is $(-++ \dots)$, the convention for the Riemann tensor 
is $\nabla_{[a} \nabla_{b]} k_c = (1/2) {R_{abc}}^d k_d$ 
and $R_{ab} = {R_{acb}}^c$ for the Ricci tensor. 
Indices in parenthesis are symmetrized and indices 
in brackets are antisymmetrized. 

\section{Asymptotic flatness in $d$ dimensions}
\label{sect:2}  

Asymptotic conditions in field theory require the specification of 
a background configuration and the precise rate at which this background 
is approached. In the case of asymptotic flatness in higher dimensional
general relativity, the background is the Minkowski metric\footnote{Other
asymptotic conditions would of course require a different background.}. 
In order to 
specify the precise rate at which Minkowski spacetime is approached at 
null infinity, it is of great technical advantage to work within a 
framework in which ``infinity'' is attached as additional points to the 
spacetime manifold, $\tilde M$ (thereby obtaining an ``unphysical'' 
spacetime manifold $M$), and in which these points are brought
metrically to a finite distance by rescaling the physical metric, $\tg_{ab}$,
by a conformal factor $\Omega^2$ with suitable properties. The asymptotic
flatness conditions are then formulated in terms of this rescaled ``unphysical
metric,''
\begin{equation}
g_{ab} = \Omega^2 \tg_{ab}, 
\end{equation}
and its relation to the likewise conformally rescaled version of 
Minkowski spacetime, 
\begin{equation}
\label{bargabom}
\bar g_{ab} = \Omega^2 \tilde \eta_{ab}.  
\end{equation}
We will refer to $(\bar M, \bar g_{ab})$ as the ``background geometry''.
As it is well-known, eq.~\eqref{bargabom} can be realized e.g. 
by conformally embedding Minkowski spacetime into a patch 
of the Einstein static universe ${\mathbb R} \times S^{d-1}$ 
with line element ${\rm d} \bar s^2 = -{\rm d} T^2 + {\rm d}\psi^2 + 
\sin^2 \psi \,{\rm d}\sigma^2$. Here, ${\rm d} \sigma^2$ is the line element 
of the unit round $(d-2)$-dimensional sphere, and  
$\psi$ is the azimuthal angle of $S^{d-1}$. In these coordinates, 
Minkowski spacetime corresponds to the region 
$\bar M = \{-\pi < T\pm \psi < \pi, \psi > 0\}$ of 
${\mathbb R} \times S^{d-1}$, see appendix~B for further details 
of this conformal embedding, as well as the specific form of $\Omega$.

The conformal infinity of Minkowski spacetime is 
the boundary of the region $\bar M$ in the Einstein static universe. 
It is divided into the five parts (1) future timelike infinity 
(the point $T=\pi, \psi=0$) (2) past timelike 
infinity (the point $T=- \pi, \psi = 0$) 
(3) spacelike infinity (the point $\psi = \pi, T= 0$) 
(4) past null-infinity (the points $T= -\pi + \psi$ for 
$0 < \psi < \pi$) and (5) future null infinity 
(the points $ T= \pi - \psi$ for $0 < \psi < \pi$). 
The conformal factor $\Omega$ is smooth in a neighborhood of null infinity
and vanishes there, and the gradient if $\Omega$ is null
there.  

Our definition of asymptotic flatness consists in specifying the precise 
rate at which $g_{ab}$ approaches $\bar g_{ab}$ as one approaches the 
boundary $\partial \bar M$. To quantify how various tensor behave 
at that boundary, we introduce the following notion:
We will say that a tensor field, $L_{ab \dots c}$, is 
``of order $\Omega^s$'' with $s \in {\mathbb R}$, written 
$L_{ab \dots c} = O(\Omega^s)$, if the tensor field 
$\Omega^{-s} L_{ab\dots c}$ is smooth at the boundary. 
It is a consequence of this definition that if $L_{ab\dots c}$ is 
of order $s$ then $\Omega^r L_{ab\dots c}$ is of order $s+r$, and 
that $\bar \nabla_{d_1} \cdots \bar \nabla_{d_k} L_{ab \dots c}$ 
is of order $s-k$. 

\medskip

We now state our definition of asymptotic flatness in even spacetime 
dimensions $d>4$. 
(From now on, $d$ is taken to be an even number unless 
stated otherwise.) 
Let $(\bar M, \bar g_{ab})$ be the background geometry described
above. A $d$-dimensional spacetime $(\tilde M, \tg_{ab})$ 
will be said to be {\it weakly asymptotically simple 
at null infinity} if the following is true: 
\begin{enumerate}
\item
It is possible to attach a boundary, $\I$, to $\tilde M$ such that 
any open neighborhood of $\I$ in $M = \tilde M \cup \I$ 
is diffeomorphic to an open subset of the manifold $\bar M$ of our 
background geometry, and such that points of $\I$ get mapped to 
(possibly a subset of) the boundary of ${\bar M}$ under this identification. 
 
\item
One has, relative to our  background metric $\bar g_{ab}$, that 
\begin{eqnarray}
\label{1a}
 {\bar g_{ab}} - g_{ab} &=& O({\Omega^{\frac{d-2}{2}}}), \quad 
 {\bar \epsilon_{ab \dots c}} - 
\epsilon_{ab \dots c} = O({\Omega^{\frac{d}{2}}}), 
\end{eqnarray}
where $\bar \epsilon_{ab \dots c}$ and $\epsilon_{ab \dots c}$ denote 
the volume element (viewed as $d$-forms) associated with the metrics 
$\bar g_{ab}$ respectively $g_{ab}$, as well as
\begin{eqnarray}
\label{1b}  
 ({\bar g^{ab}} - g^{ab}) ({\rm d} \Omega)_a = O({\Omega^{\frac{d}{2}}}),
 \quad ({\bar g^{ab}} - g^{ab})({\rm d} \Omega)_a({\rm d} \Omega)_b 
  = O({\Omega^{\frac{d+2}{2}}}), 
\end{eqnarray}
where $g^{ab}$ is the inverse of $g_{ab}$ and where 
$\bar g^{ab}$ is the inverse of $\bar g_{ab}$. 
\end{enumerate}

\smallskip 

It is important to note that, while our definition 
of an asymptotically flat spacetime is formulated relative 
to a specific background geometry, our definition is actually
independent of the precise way in which the Minkowski metric 
$\tilde \eta_{ab}$ is written as $\Omega^{-2} \bar g_{ab}$ in terms 
of a background metric (smooth at $\I$), and correspondingly the way 
in which the physical metric $\tilde g_{ab}$ is written as 
$\Omega^{-2} g_{ab}$ in terms of an auxilary unphysical metric. 
In other words, if $k$ is a smooth function defined in a neighborhood 
of the boundary of $\bar M$ such that $k \neq 0$ at null infinity, 
then our definition of an asymptotically flat metric is unchanged 
if we change the conformal factor to $\Omega' = k \Omega$, 
the background metric to $\bar g'{}_{ab} = k^2 \bar g_{ab}$, 
and the unphysical metric to $g'{}_{ab} = k^2 g_{ab}$.  

As in 4 spacetime dimensions, the notion of weak 
asymptotic simplicity can be strengthened by requiring in addition 
that every inextendible null geodesic in $(\tilde M, \tg_{ab})$ has 
precisely two endpoints on $\I$. Such a spacetime 
is then simply called {\em asymptotically simple}. 
This additional condition, combined with the fact that $\I$ is null, 
makes it possible to divide $\I$ into disjoint sets, $\I^+$ and
$\I^-$, on which future respectively past directed null geodesics 
have their endpoints. These sets are referred to as future
respectively past null infinity. This condition also implies 
that $(\tilde M, \tg_{ab})$ necessarily has 
to be globally hyperbolic, by a straightforward generalization 
of Prop.~6.9.2 of~\cite{he} to $d$ dimensions\footnote{
We also note that, by a straightforward generalization 
of Prop.~6.9.4 of~\cite{he} to $d$ dimensions, the additional
condition is in fact only consistent with $\I$ having 
topology ${\mathbb R} \times S^{d-2}$. This agrees with the topology of 
the boundary of our background geometry.}.  

Item (1) of our definition is essentially the statement that, as
manifold, $M$ looks near $\I$ like the background manifold $\bar M$ 
looks near its null boundary. Item (2) of the definition involves 
three different metrics: The physical $\tg_{ab}$, unphysical $g_{ab}$,
and the  background metric $\bar g_{ab}$. 
The physical and unphysical metric are related by the conformal
factor, $\Omega$, which makes the unphysical metric smooth and 
at the same time brings null infinity, $\I$, to a ``finite location'', 
and the background metric is likewise related to Minkowski spacetime. 
The relation between the unphysical and background metric is given 
by the above set of equations in item (2), which specify the precise 
manner in which the unphysical metric approaches the background
metric, and thus the precise sense in which our spacetime is 
required to flatten out at null infinity. Since $n_a$ is null 
relative to the background metric, it is also null relative 
to the metric $g_{ab}$, showing that $\I$ is a null surface 
in the unphysical spacetime $(M, g_{ab})$. Since $\Omega^2$ times 
Minkowski spacetime is isometric to $({\bar M}, {\bar g_{ab}})$, our 
definition of asymptotic flatness trivially covers Minkowski
spacetime. 

Since we will be working with different 
metrics in this paper---physical and unphysical---it is necessary to 
specify a rule for raising and lowering indices of tensors. 
Our rule is that indices on tensor fields on $M$ without 
a ``tilde'' are 
raised and lowered with the unphysical metric, $g_{ab}$ and 
its inverse, whereas indices on tensor fields on $\tilde M$ 
with a ``tilde'' 
are raised and lowered with the physical metric, $\tg_{ab}$, 
and its inverse\footnote{Note that this rule is consistent 
with our notation $g^{ab}$ and $\tg^{ab}$ for the inverse 
of the metrics $g_{ab}$ and $\tg_{ab}$.}. 

Let us compare the above definition of asymptotic flatness with the behavior of the $d$-dimensional
analog of the Schwarzschild metric, given by the line element
\begin{equation}
\label{ds}
{\rm d}\tilde s^2 = - \left(1 - cr^{-(d-3)} \right) {\rm d}t^2 
+ \left(1 - cr^{-(d-3)} \right)^{-1} {\rm d}r^2 + r^2 {\rm d}\sigma^2, \quad c > 0,  
\end{equation}
where ${\rm d} \sigma^2$ 
is the line element of a round $(d-2)$-dimensional sphere.
Introducing a coordinate $u$ by the relation
${\rm d}u = {\rm d}t - (1 - cr^{-(d-3)})^{-1} {\rm d}r$, the line element takes the form
\begin{equation}
\label{ds'}
{\rm d} \tilde s^2 = -2{\rm d}u{\rm d}r - {\rm d}u^2 + r^2 {\rm d}\sigma^2 + cr^{-(d-3)}{\rm d}u^2,   
\end{equation}
where the first three terms on the right side are recognized as the Minkowski line element.
Multiplying by our conformal factor $\Omega^2$, using $r^{-1} = O(\Omega)$, and using
that $\Omega^2$ times the Minkowski metric is equal to our background metric ${\rm d} \bar s^2$ by construction, it follows
that the unphysical Schwarzschild metric can be written as 
${\rm d} s^2 = {\rm d} \bar s^2 + O(\Omega^{d-1}){\rm d}u^2$ 
(noting that $u$ is a good coordinate at infinity). 
It follows that Schwarzschild spacetime is asymptotically 
flat in the sense of our definition, but it becomes flat at 
null infinity at a faster rate than that specified above 
in eqs.~\eqref{1a} and~\eqref{1b} in $d>4$. 
(In $d=4$, the relevant components drop off at the same rate, 
as specified in eqs.~\eqref{1a} and~\eqref{1b}.) 

\medskip

The above definition of asymptotic flatness in even dimensions $d>4$ is not 
appropriate in odd spacetime dimension, since condition~\eqref{1a} in item 2 now says that the unphysical metric 
$g_{ab}$ differs from the smooth background metric $\bar g_{ab}$ by a half odd integer power of $\Omega$, 
and thereby manifestly contradicts the assumption in item 1 that $g_{ab}$ is smooth at the boundary. The 
powers of $\Omega$ appearing in eqs.~\eqref{1a} and~\eqref{1b} reflect the drop off behaviour of a linearized
perturbation (see section~\ref{sect:3}), and it is hard to see how these powers could be essentially different from the ones  
in the full nonlinear theory. It therefore appears that the unphysical metric is generically at most $(d-3)/2$ times
differentiable at the boundary in odd dimensions. We note that it is also inconsistent in odd dimensions to postulate that 
the quantity $\Omega^{-(d-2)/2}(g_{ab} - \bar g_{ab})$ is smooth at the boundary as we did above in eq.~\eqref{1a} of 
item 2 in the even dimensional case, because the unphysical Schwarzschild metric $g_{ab}$ differs from the background
$\bar g_{ab}$ by terms of order $\Omega^{d-1}$, i.e., by an even power of $\Omega$. Therefore, eq.~\eqref{1a} is definitely
false for the Schwarzschild metric in odd dimensions. For the Schwarzschild metric, $\Omega^{-(d-1)}(g_{ab} - \bar g_{ab})$ 
is smooth at the boundary (in even and odd dimensions), so one might be tempted to try this condition, together with suitable 
other conditions, as the definition of asymptotic flatness. However, this would eliminate from consideration all radiating
spacetimes and is therefore not acceptable. One may try to bypass these problems by requiring appropriate 
lower differentiability properties of the corresponding quantities, but these seem neither to lead to a definition 
of asymptotic flatness that is stable under perturbations, 
as we briefly discuss in section~\ref{sect:3}, nor do those weaker conditions seem 
to be able to guarantee the 
existence of conserved quantities such as Bondi energy. Thus, it seems that a sensible definition of asymptotic simplicity 
at conformal infinity in odd spacetime dimensions would have to differ substantially from the one given above for even dimensions, 
and it is doubtful that such a definition can be cast into the framework of conformal infinity. 
For the rest of this paper, we will restrict attention to even spacetime dimensions. 

We finally comment on how the above definition of asymptotic flatness 
in even spacetime dimensions $d>4$ compares to the usual definition~\cite{g} 
in 4 dimensions. In this definition, one simply demands that there 
exists {\em some} conformal factor, $\Omega$, such that the corresponding
unphysical metric is smooth at $\I$ and such that $n_a$ is non-vanishing and 
null there\footnote{The nullness of $n_a$ follows from the first 
condition if Einstein's equations with vanishing stress energy 
at null infinity are assumed.}. 
This definition is different in appearance from that given above
and avoids in particular the introduction of a background geometry. Nevertheless, 
the definition of asymptotic flatness in $d=4$ as just stated can be brought\footnote{We emphasize, 
however, that an analogous statement is not true in $d>4$. Namely, it is not true that 
our definition of asymptotic flatness in higher dimensions is equivalent to the 
statement that there 
exists some conformal factor, $\Omega$, such that the corresponding
unphysical metric is smooth at $\I$ and such that $n_a$ is non-vanishing and 
null there.}
into a form that is very similar (but not identical) to the one given above for $d>4$. 
To see this in more detail, we recall that the usual definition of 
asymptotic flatness in 4 dimensions is equivalent~\cite{tamwi} to the statement that 
the physical metric can be cast into ``Bondi form''\footnote{
\label{footnoteee}
It is assumed in the 
derivation of eq.~\eqref{bondiform} that 
the vacuum Einstein equations are satisfied.} (see eqs.(14) and~(31)--(34) of 
\cite{bms}), 
\begin{eqnarray}
\label{bondiform}
{\rm d}\tilde s^2 &=& -2{\rm d}u{\rm d}r - {\rm d}u^2 + r^2 {\rm d} \sigma^2 \nonumber\\
           &&+ O(r) {\rm d}({\rm angles})^2 
             + O(1) {\rm d} u {\rm d} ({\rm angles}) + O(r^{-1}) {\rm d} u^2 + O(r^{-2}){\rm d} u {\rm d} r
\end{eqnarray}
in suitable coordinates near null infinity, where the first line is recognized as the Minkowski line element, 
and where ``angles'' stands for the usual polar angles of $S^2$. In $d>4$ spacetime dimensions
our asymptotic flatness conditions eqs.~\eqref{1a} and~\eqref{1b} in effect state that the physical line 
element can be written in the form
\bena
\label{bondiformd}
{\rm d}\tilde s^2 &=& -2{\rm d}u{\rm d}r - {\rm d}u^2 + r^2 {\rm d} \sigma^2 \nonumber\\
           &&+ O(r^{-\frac{d-4}{2}}) {\rm d}({\rm angles})^2 
             + O(r^{-\frac{d-4}{2}}) {\rm d} u {\rm d} ({\rm angles}) \nonumber \\
           &&+ O(r^{-\frac{d-2}{2}}) {\rm d} u^2 + O(r^{-\frac{d}{2}}){\rm d} u {\rm d} r,  
\eena
where ``angles'' now stands for the polar angles of $S^{d-2}$. 
One notices that the Bondi form~\eqref{bondiformd} in $d>4$ does not reduce to eq.~\eqref{bondiform} when $d$ is set to 4. 
The difference between the two expression arises from the ${\rm d}({\rm angles})^2$-term, which quantifies
the perturbations in the size of the cross sections of a lightcone relative to Minkowski spacetime. 
According to eq.~\eqref{bondiform}, this term is of order $O(1)$ in $d=4$ for a radiating metric, whereas
eq.~\eqref{bondiformd} would say that it ought to be of 
order $O(r^{-1})$. The latter is simply wrong for a radiating 
metric in 4 dimensions. This difference can be traced back 
to the last of conditions~\eqref{1a} in $d>4$ dimensions, 
which therefore does not hold in $d=4$. This special feature 
of $4$ dimensions will be reflected in corresponding differences 
in our discussion of the Bondi energy in dimensions $d>4$. 
We will therefore, for the rest of this paper, keep the case 
$d=4$ separate and assume throughout that $d>4$ (and even). 
Our formulas will not be valid in $d=4$ unless stated otherwise. 

\medskip

A diffeomorphism $\phi$ such that $\phi^* \tilde g_{ab}$ is 
asymtpotically flat whenever $\tilde g_{ab}$ is asymptotically flat is called an {\em asymptotic symmetry}. 
It is clear that the asymptotic symmetries form a group 
under the composition of two diffeomorphisms. Clearly, the property of being
an asymptotic symmetry is only related to the behavior of $\phi$ near the 
boundary.
An infinitesimal asymptotic symmetry is a smooth vector 
field $\xi^a$ on $\tilde M$ that has a smooth extension (denoted by the same symbol) to 
the unphysical manifold, $M$, and which generates a 1-parameter 
group of asymptotic symmetries. It is a direct consequence of our definitions
that the quantity 
\ben
\chi_{ab} = 
 \Omega^{-\frac{d-6}{2}} \pounds_{\xi} \tg_{ab} = 2\Omega^{-\frac{d-2}{2}}(\nabla_{(a} \xi_{b)} - 
\Omega^{-1} n^c \xi_c g_{ab})
\label{chiab}
\een
then has to satisfy
\begin{equation}
\label{chicond}
\chi_{ab} = O(1), \quad \chi_a{}^a = O(\Omega), \quad \chi_{ab} n^a 
= O(\Omega), \quad \chi_{ab} n^a n^b 
= O(\Omega^2), \quad \xi^a n_a = O(\Omega),   
\end{equation}
where here and in the following we are using the abbreviation 
\begin{equation}
    n_a = \nabla_a \Omega . 
\end{equation}
Conversely, if the above relations are satisfied 
for {\em some} asymptotically flat spacetime, then $\xi^a$ is 
an infinitesimal asymptotic symmetry. The classification of asymptotic 
symmetries in $d>4$ differs from that in 4 dimensions. 
We will discuss this issue in some detail below in section~5 
and in appendix~C, as well as in a fourthcoming paper~\cite{ah2}.

\section{Stability of asymptotic flatness to linear order}
\label{sect:3}

In this section we justify the definition of asymptotic 
flatness for even $d$ given in the previous section by showing that 
it is stable under linear perturbations. 
What we mean by this is the following. 
Suppose that $(\tilde M, \tg_{ab})$ is an asymptotically flat spacetime
that is also a solution to the vacuum Einstein equation, 
$\tilde R_{ab} = 0$. Consider a solution, $\delta \tg_{ab}$,  
to the linearized equations of motion around this 
background\footnote{ 
In this section, by ``background'' we mean the ``unperturbed''   
physical spacetime $(\tilde{M},\tilde{g}_{ab})$, 
unless otherwise stated.   
This should not be confused with the background~(\ref{bargabom}), 
which is our reference spacetime for defining asymptotic flatness. 
}, 
\begin{equation}
\label{dr}
0 = \delta \tilde R_{ab} 
  = \frac{1}{2}(-\tnabla^m \tnabla_m \delta \tg_{ab} 
  - \tnabla_a \tnabla_b \delta \tg_m{}^m 
  + 2\tnabla^m \tnabla_{(a} \delta \tg_{b)m}),
\end{equation}
which has the property that the restriction of $\delta \tg_{ab}$ 
to a Cauchy surface has compact support. We will show in this section 
that there exists a gauge (the transverse traceless gauge works) 
such that, setting $\delta g_{ab} = \Omega^2 \delta \tg_{ab}$,  
\begin{equation}
\label{dropoff}
\delta g_{ab} = O(\Omega^{\frac{d-2}{2}}), \quad 
\delta g_{ab} n^a = O(\Omega^{\frac{d}{2}}), 
\quad \delta g_{ab} n^a n^b 
            = O(\Omega^{\frac{d+2}{2}}), \quad 
            g^{ab} \delta g_{ab} = O(\Omega^{\frac{d}{2}}), 
\end{equation}
at $\I$ for all even $d>4$. 
These conditions are recognized as the linearized version 
of our definition of asymptotic flatness, eqs.~\eqref{1a}
and~\eqref{1b}, about an asymptotically flat background. Our definition of 
asymptotic flatness is therefore stable to linear order. 

We have emphasized above that our decay properties of the metric 
perturbations are valid only in a particular gauge. Indeed, 
since the linearized equations of motion, $\delta \tilde R_{ab}=0$, 
are invariant under a gauge transformation $\delta \tg_{ab} 
\to \delta \tg_{ab} + \pounds_{\xi} \tilde g_{ab}$ 
with $\xi^a$ an arbitrary smooth vector field on $\tilde M$, 
eq.~\eqref{dropoff} cannot possibly be true in an arbitrary gauge. For 
if it were to hold in one gauge, it would certainly not hold 
in a gauge with a $\xi^a$ that is very badly behaved at $\I$. 
Thus, the specification of an admissible 
gauge choice for the metric perturbation is an important part of 
the demonstration of eq.~\eqref{dropoff}. 

A proof that asymptotic flatness is stable to linear order 
in 4 dimensions was given by Geroch and Xanthopoulos~\cite{gx}. 
Their argument consists of the following two steps: 
One first writes the linearized equations of motions~\eqref{dr} 
in terms of the unphysical metric and derivative operator 
and introduces new field variables such that eq.~\eqref{dr} 
is transformed into a hyperbolic system of partial differential 
equations whose coefficients 
are either manifestly regular functions as one approaches $\I$
or can be made to vanish by a suitable gauge choice. 
One then argues, using standard existence and uniqueness results for solutions of 
hyperbolic partial differential equations, 
that the new variables therefore have a smooth extension 
to the unphysical spacetime. (Here one needs to use that 
the perturbation has compactly supported initial data.) 
Translating this statement about the new variables back 
into a statement about the metric perturbation, one finds 
the decay properties of the metric perturbation at $\I$. 

We here use this basic strategy to analyze the decay at $\I$ of 
metric perturbations in even dimensions $d>4$. 
The second step does not depend on the dimension of the spacetime, 
since it only involves general properties of hyperbolic differential 
equations. By contrast, the first step, i.e. the actual choice of 
variables and gauge conditions, is different in nature in $d > 4$ 
dimensions as compared to $d=4$. Finding the appropriate variables 
and gauge conditions that do the job is, of course, 
the hard part of the analysis. It needs to be done before writing down any 
decay properties of the perturbations, which are then supposed 
to follow from the precise form of the new variables and gauge
conditions. We here present things in the opposite order 
in order to simplify the exposition. (Throughout the rest of 
this section, $d>4$ and even will be assumed.) 
 
Concerning the proper choice of gauge, we consider the 1-form
\begin{equation}
\label{vadef}
\tilde v_a = \tg^{bc} \tnabla_c \delta \tg_{ab} 
           - \tg^{bc} \tnabla_a \delta \tg_{bc}, 
\end{equation}
which is equal to (dual of) the integrand of the surface 
term arising when varying the Einstein-Hilbert action. Under a gauge 
transformation, $\delta \tg_{ab} \to \delta \tg_{ab} 
+ \pounds_\xi \tg_{ab}$, this quantity is seen to transform as 
\begin{equation}
\label{vatrans}
\tilde v^a \to \tilde v^a + 2\tnabla_b \tnabla^{[b} \xi^{a]} ,  
\end{equation}
where the vacuum Einstein equations $\tilde R_{ab}=0$ have been used. 
On the other hand, the variation of the Ricci tensor can be seen to satisfy
\begin{equation}
\tg^{ab} \delta \tilde R_{ab} = \tnabla^a \tilde v_a, 
\end{equation}
so that we have $\tnabla^a \tilde v_a = 0$
when the linearized Einstein equations hold. 
Thus, by eq.~\eqref{vatrans}, we can set 
$\tilde v_a = 0$ throughout $\tilde M$ 
by a choice of gauge transformation when the linearized Einstein 
equations hold\footnote{
Note that $0=\tilde v^a + 2\tnabla_b \tnabla^{[b} \xi^{a]}$ 
has the same form as Maxwell's equation for a vector potential, 
with a divergence free
source. A solution therefore exists by the same arguments 
as for Maxwell's equations.}. 
This gauge choice is invariant under further gauge transformations 
of the form $\delta \tg_{ab} \to \delta \tg_{ab} + \pounds_\xi \tg_{ab}$, 
with $\xi^a = \tnabla^a \xi$ and $\xi$ any smooth function on $\tilde M$, 
which can be used, for example, 
to impose a gauge condition on the trace of the perturbation. We impose 
\begin{equation}
\label{gc}
\delta \tg_a{}^a = 0, 
\end{equation}
which can always be realized since the trace transforms as 
\begin{equation}
\delta \tg_a{}^a \to \delta \tg_a{}^a + \tnabla^a \tnabla_a \xi
\end{equation}
under the remaining gauge transformations. Thus, our gauge conditions are eq.~\eqref{gc} and $\tilde v_a = 0$. Together, 
they are equivalent to the transverse traceless gauge condition, $\tnabla^a \delta \tg_{ab} = \delta \tg_m{}^m = 0$.

Concerning the proper choice of field variables when $d>4$, we consider
\begin{equation}
\label{vardef}
\tau_{ab} \equiv \Omega^{-\frac{d-2}{2}} \delta g_{ab}, \quad \tau_a \equiv 
\Omega^{-1} \tau_{ab} n^b, \quad u \equiv \nabla^a \tau_a. 
\end{equation}
We substitute the definitions~\eqref{vardef} into the linearized Einstein equation, and use the background 
Einstein equation, $\tilde R_{ab} = 0$, as well as the well-known relations between the physical 
and unphysical derivative operator and Ricci tensor, 
\begin{equation}
\label{der}
\tnabla_a k_b = \nabla_a k_b + \Omega^{-1}(2 {\delta^c}_{(a} n_{b)} - g_{ab} n^c) k_c, 
\end{equation}
\begin{equation}
\label{ric}
\tilde R_{ab} = R_{ab} + \Omega^{-1}[(d-2)\nabla_a n_b + (\nabla_m n^m) g_{ab}
- (d-1) fg_{ab}].
\end{equation}
A lengthy calculation shows that the result can be written in the form
\begin{eqnarray}
\label{lee}
 0 = 2 \Omega^{-\frac{d-2}{2}} \delta \tilde R_{ab} 
      &=& 
           - \nabla^c \nabla_c \tau_{ab}
           + \nabla_a \nabla_b \tau 
           + 4 \nabla_{(a} \tau_{b)} 
           + 2 \nabla_{(a} y_{b)}
           - 2ug_{ab}  
 \non \\
       && \, 
           - 2 R_{acbd}\tau^{cd} 
           - \frac{(d-6)}{2(d-2)} R_{ab} \tau      
           + \frac{(d-2)}{4(d-1)} R \tau_{ab}
           + \frac{(d-6)}{4(d-1)(d-2)} g_{ab} R \tau       
\non \\ 
       && \, 
       + \Omega^{-1} 
             (d-2) n_{(a} y_{b)} 
       + \Omega^{-1} g_{ab} \left( 
                           (d-2)n_c\tau^c + n^c\nabla_c\tau 
                           + \frac{(3d-10)}{4}f\tau 
                          \right)
\non \\ 
       && \, 
           + (d-4)\Omega^{-1} n_{(a} \nabla_{b)} \tau        
           + \frac{(d-6)(d-4)}{4}\Omega^{-2} n_an_b \tau
        \,,           
\end{eqnarray}
where we have used the shorthand notation
\begin{equation}
\label{yadef}
  \tau = {\tau^a}_a, \quad y_a = \Omega^{-\frac{d-2}{2}} \tilde v_a = 
  \nabla^c\tau_{ca} - \nabla_a \tau - \f((d+2)/2) \tau_a 
                            - \f((d-4)/2) \Omega^{-1}n_a \tau,  
\end{equation}
as well as 
\begin{equation}
f = \Omega^{-1} n_a n^a. 
\end{equation}
We now substitute our gauge conditions, $y_a = \tau = 0$, using in particular that
\begin{eqnarray}
\label{naya}
0 = n^a y_a &=&\frac{1}{2} \left(u + \frac{1}{(d-2)} R^{ab} \tau_{ab} \right)
- \frac{d}{4} \Omega^{-1} n^a \tau_a,  
\end{eqnarray}
which follows by combining our gauge condition with the background Einstein equation, to
get rid of the $\Omega^{-1} n^a \tau_a$ term in eq.~\eqref{lee}.
Then eq.~\eqref{lee} reduces to  
\begin{equation}
\label{tab}
\nabla^c \nabla_c \tau_{ab} =
4\nabla_{(a} \tau_{b)} - \frac{4}{d} u g_{ab} 
+ \frac{(d-2)}{4(d-1)} R \tau_{ab} + \frac{2}{d}R^{cd} \tau_{cd} g_{ab} 
- 2R_{acbd} \tau^{cd},   
\end{equation}
where all singular terms now have dropped out due to our choice of variables and 
gauge condition. We are, however, not done yet since we also need appropriate equations 
for the remaining variables, $\tau_a, u$.

In order to get an equation for $\tau_a$, 
we take the divergence of eq.~\eqref{tab} with respect to the 
unphysical metric and use again the transverse traceless 
gauge condition, $y_a = \tau = 0$. This gives 
\begin{eqnarray}
\label{eq2}
(d-2)\nabla^c \nabla_c \tau_{a} &=&  
\frac{4(d-2)}{d}\nabla_{a} u + \frac{4}{d} R^{bc}\nabla_a \tau_{bc} 
   - 2R^{bc} \nabla_{b} \tau_{ca}\nonumber \\  
&& + \left(\frac{(d+2)(d-2)}{4(d-1)} R g_{ab} + 4R_{ab} \right) \tau^b 
   \nonumber\\ 
&&
 + 2\left(
          \nabla_b R_{ca} - \frac{(d-2)}{d}\nabla_a R_{bc}  - \frac{d}{4(d-1)} (\nabla_c R)g_{ab} 
    \right) \tau^{bc} .
\end{eqnarray}
Finally, in order to get an equation for $u$, we take a further divergence of eq.~\eqref{eq2}. We 
use eq.~\eqref{tab} and the divergence of the transverse traceless gauge condition $\nabla^a y_a = \tau = 0$
to eliminate second derivatives of $\tau_{ab}$, giving
\begin{eqnarray}
\label{eq3}
(d-2)(d-4) \nabla^c \nabla_c u
      &=&
       - 2(d+2)(d-4)R_{ab} \nabla^a \tau^b
       - 2(d-4)\nabla^a R^{bc}\nabla_a \tau_{bc}
 \nonumber \\
      &&
      + \frac{(d-4)(d^3+4d^2+12d-16)}{4d(d-1)} R u 
      - \frac{d(d-2)(d-4)}{2(d-1)} (\nabla_c R) \tau^c 
\nonumber \\ 
     && 
      + \left(
             \frac{d(d-2)}{2(d-1)} \nabla_a \nabla_b R 
            - 2(d-2) \nabla^c\nabla_c R_{ab}  
        \right.
\nonumber \\ 
     && 
        \qquad 
        \left.
            - 8R^{cd} R_{acbd} 
            + \frac{d^2+6d-8}{d(d-1)} R R_{ab} 
        \right) \tau^{ab} \,. 
%
\end{eqnarray}
Equations~\eqref{tab}, \eqref{eq2} and~\eqref{eq3} form a system of linear partial differential equations 
for the variables $\tau_{ab}, \tau_a, u$ in the unphysical spacetime $M$, with coefficients that are 
given in terms of the unphysical Riemann and Ricci tensor and its first and second 
derivatives. No terms containing explicitly inverse powers of $\Omega$ appear due to our particular choice of variables 
and gauge conditions. Introducing the shorthand notation $\phi_\alpha = (\tau_{ab}, \tau_a, u)$, this 
system can be rewritten more compactly as
\begin{equation}
\label{pde}
g^{ab} \nabla_a \nabla_b \phi_\alpha = A_\alpha{}^{\beta a} \nabla_a \phi_\beta + B_\alpha{}^\beta \phi_\beta. 
\end{equation}
It follows from our definition of an asymptotically flat spacetime that the coefficients $A_\alpha{}^{\beta a}, 
B_\alpha{}^{\beta}$ in this system are smooth tensor fields up to and on the boundary. 
Equation~\eqref{pde} therefore forms a hyperbolic system\footnote{I.e., 
roughly speaking, does not contain any second derivatives other than the wave operator. Such terms 
could have arisen via expressions such as $R^{abcd} \nabla_a \nabla_c \tau_{bd}$ which, as we note,
could not be eliminated in favor of first derivative
terms via our gauge condition, $\nabla^m y_m = 0$. Fortunately, these terms happen to cancel.} 
of partial differential equations with coefficients that are smooth functions up to and on $\I$. 
Hence, this system possesses a well-posed initial value formulation~\cite{he} in the unphysical 
spacetime. If $\phi_\alpha$ has compactly supported initial data as we have assumed, then we conclude
by the general argument given in~\cite{gx} that $\phi_\alpha$ and hence $\tau_{ab}, \tau_a$ and $u$, extend to 
smooth tensor fields at $\I$. In combination with eq.~\eqref{naya}, this implies moreover
that $\Omega^{-1} n^a \tau_a$ is smooth at $\I$ in the transverse traceless gauge, and hence that 
$n^a \tau_a = O(\Omega)$. Substituting back the 
definition~\eqref{vardef} of $\tau_{ab}$ and $\tau_a$ and the gauge condition $\tau = 0$ 
in terms of $\delta g_{ab}$, we altogether find that 
the desired drop off properties~\eqref{dropoff} hold for 
the linearized perturbation at $\I$. Thus, we have shown 
that our definition of asymptotic simplicity given in the previous section is stable 
under linear perturbations when $d>4$ and even.

\medskip

For completeness, we now comment upon the status of the above argument in the case when $d>4$ and odd. In that case, 
the algebra leading to eq.~\eqref{pde} is identical as in the case $d>4$ and even, but the coefficients $g^{ab}, A_\alpha{}^{\beta a}, 
B_\alpha{}^{\beta}$ in this system now cannot be assumed to be smooth at the boundary, since the unphysical metric
$g_{ab}$ itself does not have this property (see the discussion in
section~\ref{sect:2}). 
Instead, since $g_{ab}$ can at best 
be expected to be of differentiability class $C^s, s=(d-2)/2$ at the boundary, 
we can at best expect that $A_\alpha{}^{\beta a} \in C^{s-3}, B_\alpha{}^{\beta} \in C^{s-4}$ at the boundary. 
On the other hand, the standard existence and uniqueness results for linear hyperbolic equations of the form~\eqref{pde}
require a higher degree of regularity\footnote{In order to guarantee existence and uniqueness of a solution in the class
$\phi_\alpha \in W^{d/2 + 2 + A}, A \ge 0$ (we mean the Sobolev space), one needs $g_{ab} \in W^{d/2 + 2 + A}$, 
$A_\alpha{}^{\beta a} \in W^{d/2 + 1 + A}$ and $B_\alpha{}^\beta \in W^{d/2 + 1 + A}$. This is stronger than what we know.}
for the coefficients and therefore do not 
guarantee the existence of a solution to~\eqref{pde}. Thus, unlike in the case of even $d>4$, we now cannot conclude 
that $\phi_\alpha$ and hence $\tau_{ab}, \tau_a$ and $u$, extend to, say continuous, tensor fields at $\I$, and 
we therefore also cannot conclude that e.g. $\delta g_{ab}$ is given by $\Omega^{(d-2)/2}$ times a continuous 
function. Thus, our 
stability proof breaks down in odd dimensions. We believe that this is an indication that a geometric definition of 
asymptotic simplicity that is stable against perturbations is not possible in odd dimensions. 

\medskip

In $d=4$ spacetime dimensions, the above system of equations for $\tau_{ab}, \tau_a, u$ 
fails to be even hyperbolic, since the ``box term'' drops 
out in the equation~\eqref{eq3} for $u$. Thus, the above choice of variables and gauge does not work in $d=4$. A 
set of variables and gauge conditions that works in 4 dimensions has been found by Geroch and 
Xanthopoulos~\cite{gx}: 
These variables are $\tau_{ab}, \tau_b$ and $\sigma = 
\Omega^{-1}(n^a \nabla_a \tau + \frac{1}{2} n^a \tau_a + \frac{1}{4}f \tau)$.  
The gauge condition is chosen to be $y_a = 0$, together with a certain complicated gauge condition 
on the trace of the perturbation instead of $\tau = 0$. With this choice of variables and gauge conditions in place, 
it is then shown that $\tau_{ab}, \tau_b, \sigma$
satisfy a system of hyperbolic equations with coefficients that are smooth at $\I$ (assuming that the 
unphysical metric is smooth at $\I$). It follows now  
that the metric perturbation has the fall-off rate 
\begin{equation}
\delta g_{ab} = O(\Omega), \quad \delta g_{ab} n^b = O(\Omega^2), \quad \delta g_{ab} n^a n^b = O(\Omega^3)
\end{equation}
in this gauge. This differs from the corresponding result eq.~\eqref{dropoff} $d>4$ in that the trace of 
the perturbation is now only falling off as fast as the metric perturbation itself, $\delta g_m{}^m = O(\Omega)$,
rather than one power faster as in $d>4$. This confirms the observation already made in the previous section 
that we cannot impose the second of eqs.~\eqref{1b} in 4 dimensions, which, as we note, would be the non-linear analog
of the condition $\delta g_m{}^m = O(\Omega^2)$. Hence, it is seen that the 
definition of asymptotic flatness is qualitatively different in $d>4$ dimensions.
As we will see, this has consequences for our analysis
of the Bondi energy in $d>4$ dimensions.

\section{General strategy for defining ``conserved'' 
quantities at infinity}
\label{sect:4}

In this section, we review the general algorithm given by Wald and Zoupas~\cite{wz}
for defining ``charges'' associated with symmetries preserving
a given set of ``boundary conditions'' in the context of theories derived from a 
diffeomorphism covariant Lagrangian. This will later be used to define 
the Bondi energy in $d$-dimensional general relativity 
as the generator conjugate to an appropriately defined 
asymptotic time translation symmetry.

The algorithm~\cite{wz} applies to arbitrary theories derived from a 
diffeomorphism covariant Lagrangian. We will focus here on vacuum general relativity in 
$d$ dimensions, defined by the Lagrangian density (viewed as 
a $d$-form)
\begin{equation}
L = \frac{1}{16\pi G} \tilde R \tilde \epsilon,  
\end{equation}
and the boundary conditions specified in our definition of asymptotic flatness. 

One considers the  variation of $L$, which can always be written in the form
\begin{equation}
\delta L = E + {\rm d} \theta, 
\end{equation}
where $E$ are the equations of motion; in our case 
\begin{equation}
E_{a_1 \dots a_d} = \frac{1}{16\pi G}
                    \left( 
                       \tilde R_{bc} - \frac{1}{2}\tilde R \tg_{bc} 
                    \right) 
                       \delta \tg^{bc} \tilde \epsilon_{a_1 \dots a_d}; 
\end{equation}
and where ${\rm d}\theta$ is the exterior differential of a $(d-1)$-form $\theta$, given in our case by
\begin{equation}
\label{td}
\theta_{a_1 \dots a_{d-1}} = \frac{1}{16\pi G} \tilde v^c \tilde \epsilon_{ca_1 \dots a_{d-1}}, 
\end{equation}
where $\tilde v_a$ is given in terms of $\delta \tg_{ab}$ by eq.~\eqref{vadef}. The antisymmetrized second 
variation\footnote{Here, and in similar other formulas involving second variations, we assume without loss of 
generality that the variations 
commute, i.e., that $\delta_1(\delta_2 \tg) - \delta_2(\delta_1 \tg) = 0$.}
$\omega$ of $\theta$ defines the (dualized) symplectic current, 
\begin{equation}
\label{dw}
\omega(\tg; \delta_1 \tg, \delta_2 \tg) = \delta_1 \theta(\tg; \delta_2 \tg) - \delta_2 \theta(\tg; \delta_1 \tg),     
\end{equation}
so that $\omega$ depends on the unperturbed metric and is skew in the pair of perturbations $(\delta_1 \tg_{ab}, \delta_2 \tg_{ab})$. 
It is given in our case by 
\begin{equation}
\omega_{a_1 \dots a_{d-1}} = \frac{1}{16\pi G} \tilde w^c \tilde \epsilon_{ca_1 \dots a_{d-1}}, 
\end{equation}
where $\tilde w^c$ is the symplectic current vector
\begin{equation}
\label{wdef}
\tilde w^a = \tilde P^{abcdef}(\delta_1 \tg_{bc} \tnabla_d \delta_2 \tg_{ef} - \delta_2 \tg_{bc} \tnabla_d \delta_1 \tg_{ef})
\end{equation}
with 
\begin{equation}
\tilde P^{abcdef} 
  = \tg^{ae} \tg^{fb} \tg^{cd} 
  - \frac{1}{2} \tg^{ad} \tg^{be} \tg^{fc} 
  - \frac{1}{2}\tg^{ab} \tg^{cd} \tg^{ef} 
  - \frac{1}{2}\tg^{bc} \tg^{ae} \tg^{fd} 
  + \frac{1}{2}\tg^{bc} \tg^{ad} \tg^{ef}. 
\end{equation}
The integral of the symplectic current over an achronal 
$(d-1)$-dimensional submanifold $\tilde \Sigma$ of $\tilde M$ defines the 
symplectic structure, $\sigma$, of general relativity
\begin{equation}
\sigma(\tg; \delta_1 \tg, \delta_2 \tg) = \int_{\tilde \Sigma} \omega. 
\end{equation}
It can be shown that, when both $\delta_1 \tg_{ab}$ 
and $\delta_2 \tg_{ab}$ satisfy the linearized equations 
of motion~\eqref{dr}, then ${\rm d} \omega = 0$, or, what 
is the same thing, that the symplectic current~\eqref{wdef} 
is conserved, $\tnabla^a \tilde w_a = 0$. Consequently, 
the symplectic structure $\sigma$ 
does not depend on the choice of $\Sigma$, when $\delta_1 \tg_{ab}$ 
and $\delta_2 \tg_{ab}$ satisfy 
the linearized equations of motion with compactly supported initial data. 

The algorithm~\cite{wz} for defining generators associated with 
asymptotic symmetries now consists of the following steps. 
First, check whether the symplectic current form 
$\omega(\tg; \delta_1 \tg, \delta_2 \tg)$ has a well-defined 
(i.e., finite) extension to $\I$ for all asymptotically flat 
metrics satisfying Einstein's equation and all metric perturbations 
preserving asymptotic flatness to first order, i.e., 
for all linear perturbations satisfying the linearized equations of 
motion and eqs.~\eqref{dropoff}. 
If this is the case, one secondly seeks a $(d-1)$-form 
$\Theta(\tg; \delta \tg)$ on $\I$ which is linear in the perturbation, 
$\delta \tg_{ab}$, which is locally constructed out of 
the metric $\tg_{ab}$ and its derivatives at the boundary 
and any further quantities arising in the specification of 
the boundary condition, and which has the property that 
the pull back of the symplectic current $\omega$ to $\I$ can be 
written as the antisymmetrized variation of 
$\Theta$,  
\begin{equation}
\label{TT}
\omega(\tg; \delta_1 \tg, \delta_2 \tg) 
 = \delta_1 \Theta(\tg; \delta_2 \tg) - \delta_2 \Theta(\tg; \delta_1 \tg). 
\end{equation} 
If such a symplectic potential $\Theta$ exists (which is by no means 
guaranteed and depends crucially on the precise 
form of the boundary conditions under consideration), then 
a generator conjugate to an asymptotic symmetry can be defined as follows: 
If $\xi^a$ is a vector field
on $\tilde M$ representing an infinitesimal asymptotic symmetry,
define the associated charge ${\mathcal H}_\xi$ by the formula 
\begin{equation}
\label{db}
\delta {\mathcal H}_\xi = \int_B (\delta Q_\xi - \xi \cdot \theta) + \int_B \xi \cdot \Theta, 
\end{equation}
where $B$ is the cross section of $\I$ at which the generator is to be evaluated and where $Q_\xi$ is the 
Noether-charge $(d-2)$-form, given in the present case by 
\begin{equation}
\label{nc}
Q_{a_1 \dots a_{d-2}} = -\frac{1}{16\pi G}(\tnabla^b \xi^c) \tilde \epsilon_{bca_1 \dots a_{d-2}}.   
\end{equation}
In these formulae, the notation ``$\xi \cdot A$'' means that the vector field $\xi^a$ is contracted into the first index of a 
differential form $A$.

It is not immediately evident from what we have said so far that eq.~\eqref{db} actually defines a generator (up to an arbitrary constant), 
i.e., that the right side of eq.~\eqref{db} is indeed the ``$\delta$'' of some quantity ${\mathcal H}_\xi$. To see this, 
one first verifies that the right side of eq.~\eqref{db} has a vanishing anti-symmetrized 
second variation\footnote{This would not be so if we had not added the $\Theta$-term to the expression for 
$\delta {\mathcal H}_\xi$!}. This is certainly a necessary condition for it to arise as the  ``$\delta$'' of some quantity ${\mathcal H}_\xi$, 
for we always have $(\delta_1 \delta_2 - \delta_2 \delta_1){\mathcal
H}_\xi = 0$. 
As argued in~\cite{wz}, this is also 
a sufficient condition if one assumes that the space of asymptotically
flat metrics 
is simply connected\footnote{Note the analogy to 
``Poincare's lemma'' which says that every closed 1-form on a simply
connected space is exact, i.e., the ``${\rm d}$'' 
of some scalar function.}. 
For the cases considered in this paper, we will prove existence of 
a ${\mathcal H}_\xi$ and provide an explicit expression solving eq.~\eqref{db}.  
The arbitrary constant is fixed by setting ${\mathcal H}_\xi$ equal to 0
on Minkowski spacetime. 

The ``flux'' through a segment $\S$ of 
$\I$ bounded by two cross-sections $B_1$ and $B_2$ associated with the infinitesimal 
symmetry $\xi^a$ is defined to be the difference
\begin{equation}
\label{flux}
F_\xi = {\mathcal H}_\xi(B_2) - {\mathcal H}_\xi(B_1).  
\end{equation}
One finds the simple formula~\cite{wz}
\begin{equation}
\label{fluxf}
F_\xi = \int_{\S} \Theta(\tg; \pounds_\xi \tg). 
\end{equation}

We finally comment on the meaning of the integrals in eq.~\eqref{db}. The second integral on the right
side of~\eqref{db} has a straightforward meaning since it has been assumed that 
the integrand, $\Theta$, is well defined and smooth on $\I$. This is, however, 
not so for the first integral on the right side of~\eqref{db}, because
the integrand is defined only in the interior of the spacetime. 
This integral is to be understood instead as the limit of the corresponding
integrals for a sequence of closed, smooth $(d-2)$-surfaces $B_i$ in the interior of the physical spacetime $\tilde M$ that smoothly approach the cross section $B$ of $\I$ as $i \to \infty$. 
The following argument~\cite{wz} shows that this limit indeed exists under the assumptions that have been made: 
Let $I_i = \int_{B_i} [\delta Q_\xi - \xi \cdot \theta]$. Then, since 
\begin{equation}
{\rm d} \left[
          \delta Q_\xi(\tg, \delta \tg) - \xi \cdot \theta(\tg; \delta
          \tg) 
        \right] 
        = \omega(\tg; \delta \tg, \pounds_\xi \tg), 
\end{equation}
we have, by Stoke's theorem
\begin{equation}
I_i - I_j = \int_{\tilde \Sigma_{ij}} \omega(\tg; \delta \tg, \pounds_\xi \tg), 
\end{equation}
where $\tilde \Sigma_{ij}$ is a smooth spacelike $(d-1)$-surface bounded by $B_i$ and $B_j$. But $\omega$ has a 
smooth extension to $\I$ by assumption, so the right side of this equation goes to 0 as $i, j \to \infty$. 

\section{The Bondi energy formula}
\label{sect:5}

The aim of this section is to implement the strategy of the previous section for an asymptotic time translation 
symmetry in even spacetime dimensions $d>4$, i.e., to show that a generator ${\mathcal H}_\xi$ exists for such a symmetry, and
to derive an expression for this generator.  
We will take this ${\mathcal H}_\xi$ as the definition of the Bondi energy-momentum (Bondi energy, for short) 
in spacetimes of even dimension $d>4$. 

The crucial issue regarding the existence of a generator conjugate to asymptotic symmetries 
is whether the symplectic current $\omega$ has a (finite)
restriction to $\I$ and whether there exists, under our choice of boundary conditions, 
a potential $\Theta$ for the pull-back~\eqref{cur} of 
the symplectic current density to $\I$, i.e., a $\Theta$ satisfying~\eqref{TT}. We now examine these issues. 

We fix the conformal factor $\Omega$ once and for all, so that,  
if $(\tilde M, \tg_{ab})$ is an asymptotically flat spacetime, then $g_{ab} = \Omega^2 \tg_{ab}$ 
satisfies eqs.~\eqref{drrr} with this fixed choice of $\Omega$. Consider a solution $\delta\tg_{ab}$ of the linearized 
field equations that preserves asymptotic flatness to first order. 
Then the quantities $\tau_{ab} = \Omega^{-\frac{d-6}{2}} \delta \tg_{ab}, 
\tau_a = \Omega^{-1} \tau_{ab} n^b$ are finite and smooth at $\I$ and $\tau_a{}^a$ and $n^a \tau_a$ vanish at $\I$. 
We substitute these relations into the definition of the symplectic current, use the relation~\eqref{der}
between the physical and unphysical derivative, the relation 
\begin{equation}
\tilde \epsilon_{s_1 \dots s_d}
= \Omega^{-d} \epsilon_{s_1 \dots s_d}
\end{equation}
between the physical and unphysical volume element, 
and evaluate at $\I$. 
After some algebra, we find the simple result
\begin{equation}
\label{f}
\omega_{s_1 \dots s_d} = \frac{1}{32\pi G}
                  \left( 
                        \tau_1^{bc} \nabla^m \tau_{2 bc} 
                       - \tau_2^{bc} \nabla^m \tau_{1 bc}
                  \right) 
                         \epsilon_{ma_1 \dots a_{d-1}} + O(\Omega),    
\end{equation} 
noticing that this is finite at $\I$ (this formula is valid only in
$d>4$). This expression can be 
rewritten somewhat more conveniently introducing a $(d-1)$-form $^{(d-1)} \epsilon$ by the formula\footnote{Note
that $^{(d-1)} \epsilon$ is only defined up to the addition of a $(d-1)$-form of the form $n \wedge {^{(d-2)} \phi}$, 
where ${^{(d-2)} \phi}$ is arbitrary. The addition of such a form does however not make any difference in the formulae
given below.} 
\begin{equation}
\label{epsd-1}
{^{(d)} \epsilon_{ma_1 \dots a_{d-1}}} = d \cdot n_{[m} {^{(d-1)} \epsilon_{a_1 \dots a_{d-1}]}}  
= (n \wedge {^{(d-1)} \epsilon})_{ma_1 \dots a_{d-1}}, 
\end{equation}
where we have put a superscript on the quantities in order to indicate the degree of the 
form. The pull back to $\I$ of the symplectic current
$(d-1)$-form $\omega$ can then be written as
\begin{eqnarray}
\label{cur}
\zeta^* \omega_{a_1 \dots a_{d-1}} 
= \frac{1}{32\pi G}
  \left( 
      \tau_1^{bc} n^m \nabla_m \tau_{2 bc} 
     - \tau_2^{bc} n^m \nabla_m \tau_{1 bc}
  \right) \epsilon_{a_1 \dots a_{d-1}}, 
\end{eqnarray} 
where $\zeta^*$ denotes the pull-back of a covariant 
tensor field to $\I$. Thus, it follows from this expression 
that the pull-back of the symplectic current form to $\I$ is 
finite and smooth for any linear perturbation 
preserving our asymptotic flatness condition. 

We next look for a potential $\Theta$ for~\eqref{cur}.
The subsequent calculations are somewhat simplified using 
the tensor $S_{ab}$ defined by the equation
\begin{equation}
\label{cdefn}
R_{abcd} = C_{abcd} + g_{a[c} S_{d]b} - g_{b[c} S_{d]a} 
\end{equation}
in terms of the unphysical Riemann tensor and Weyl tensor. 
It can be expressed in terms of the unphysical Ricci tensor by
\begin{equation}
\label{sabdef}
S_{ab} = \frac{2}{(d-2)} R_{ab} - \frac{1}{(d-1)(d-2)}Rg_{ab}. 
\end{equation}
Using the relation~\eqref{ric} between the 
physical and unphysical Ricci tensor, we find that Einstein's equation
for the physical Ricci tensor takes the form 
\begin{equation}
\label{ee}
\Omega S_{ab} + 2\nabla_a n_b - fg_{ab} =0
\end{equation}
in terms of the tensor $S_{ab}$, where we remember the shorthand $f = \Omega^{-1} n^a n_a = O(1)$.
We now take the variation of eq.~\eqref{ee}
with respect to a linearized solution of the field equations 
that preserves asymptotic flatness 
to first order (remembering $\delta \Omega = 0$ since $\Omega$ 
is rigidly fixed) and substitute the definition 
of $\tau_{ab}$ and $\tau_b$. Using the formulae  
\begin{equation}
\label{delf}
\delta f = - \Omega^{\frac{d-2}{2}} \tau_a n^a, 
\end{equation}
\begin{equation}
\delta(\nabla_a n_b) 
= \Omega^{\frac{d-2}{2}} 
   \left( 
         -\frac{d}{2} n_{(a} \tau_{b)} 
         + \frac{d}{4} f\tau_{ab} 
         + \frac{1}{2} n^c \nabla_c \tau_{ab} - \Omega \nabla_{(a}\tau_{b)} 
         - \frac{1}{2} \Omega S_{c(a} {\tau_{b)}}^c 
   \right),  
\end{equation}
we find the result
\begin{multline}
\label{exp1}
\Omega^{-\frac{d-4}{2}} \delta S_{ab} 
    = d \cdot n_{(a} \tau_{b)} - n^c \nabla_c \tau_{ab} -\frac{(d-2)}{2} f  \tau_{ab}\\
      + \Omega\bigg( 2\nabla_{(a} \tau_{b)} + S_{c(a} {\tau_{b)}}^c
      -\Omega^{-1} \tau_c n^c g_{ab}
      \bigg),   
\end{multline}
and we note that the right side of this equation is manifestly finite 
at $\I$ and that the term in the second line is of order $\Omega$. 
It follows from this relation, together with the 
relation $\tau_{ab} n^b = O(\Omega)$, that the pull-back 
of the symplectic current density at $\I$, eq.~\eqref{cur}, can be written as 
\begin{equation}
\label{symplrest}
\zeta^* \omega_{a_1 \dots a_{d-1}} 
  = \frac{1}{32\pi G} \Omega^{-\frac{d-4}{2}}
    \left( 
          \tau_2^{cd} \delta_1 S_{cd} - \tau_1^{cd} \delta_2 S_{cd} 
    \right) \,\epsilon_{a_1 \dots a_{d-1}}.  
\end{equation}
To construct the desired potential, $\Theta$, for eq.~\eqref{symplrest}, we 
note that our asymptotic flatness conditions~\eqref{1a} and~\eqref{1b} imply 
that\footnote{We note that the second relation is in general false in $d=4$.} 
\begin{equation}\label{sabs}
S_{ab} - \bar S_{ab} = O(\Omega^{\frac{d-4}{2}}), \quad S_m{}^m - \bar S_m{}^m = O(\Omega^{\frac{d-2}{2}}), 
\quad f - \bar f = O(\Omega^{\frac{d}{2}}).   
\end{equation}
By construction, the restriction of the symplectic form to $\I$, eq.~\eqref{symplrest}, 
only depends on the physical metric $\tilde g_{ab}$ and its variations, 
but not on how we have chosen to write them 
in terms of an unphysical metric $g_{ab}$ and a conformal factor $\Omega$, 
as long as $g_{ab}$ and $\bar g_{ab}$ are smooth at $\I$. We now take 
advantage of this fact by choosing a conformal factor 
so that the background metric~(\ref{bargabom}) 
is flat in a neighborhood of $\I$, i.e. $\bar S_{ab} = 0$, and such that 
$\bar f = const.$ there. This is possible, at least locally in 
a neighborhood of any open subset of $\I^+$ resp. $\I^-$ with compact 
closure not intersecting spatial infinity (see appendix B for
details). Consequently, by eq.~\eqref{sabs}, in such a gauge we have 
\ben 
S_{ab} = O(\Omega^{\frac{d-4}{2}}), \quad S_m{}^m = O(\Omega^{\frac{d-2}{2}}), 
\quad f = const. + O(\Omega^{\frac{d}{2}}).   
\een 
Choose any smooth covector field $l_a$ on $M$ with the property 
that $l_a l^a = 0, n^a l_a = +1$ at $\I$, set 
\begin{equation}
q_{ab} = g_{ab} - 2n_{(a} l_{b)}
\end{equation}
and define the ``news tensor'' by 
\begin{equation}
\label{newsdef}
N_{ab} = \zeta^* \bigg(\Omega^{-\frac{d-4}{2}} q^m{}_a q^n{}_b S_{mn} \bigg) 
        \quad \text{(assuming $d>4$)},   
\end{equation}
where $\zeta^*$ denotes the pull back to $\I$. By definition, 
$N_{ab}$ is a well-defined smooth tensor field at $\I$ with vanishing 
trace. Using the identity 
\begin{equation}
\label{xx}
S_{ab} n^b + \nabla_a f = 0, 
\end{equation}
one sees that $N_{ab}$ is independent of the particular choice of $l^a$.
A symplectic potential $\Theta$ at $\I$ with the desired properties 
is now given by  
\begin{equation}
\label{Td}
\Theta_{a_1 \dots a_{d-1}} \equiv \frac{1}{32\pi G} 
\tau^{cd} N_{cd} \, \epsilon_{a_1 \dots a_{d-1}}. 
\end{equation}
While the restriction of the symplectic form $\omega$ 
to $\I$ (see eq.~\eqref{symplrest}) only depends on the physical metric and its perturbation, but 
not on the particular choice of the conformal factor $\Omega$ and unphysical metric
(although the latter were used to obtain a convenient form for $\omega$), 
this need not be the case for $\Theta$, as the latter is a potential 
for $\omega$ and therefore only unique up to a ``total variation'' $\delta \Pi$. 
We therefore have to investigate the behavior of $\Theta$ under ``conformal gauge changes''.
If we change the conformal factor to $\Omega' = k \Omega$ with some smooth $k$, 
then, since the physical metric is to remain unchanged, the unphysical metric changes
by $g'{}_{ab} = k^2 g_{ab}$, and likewise the background metric changes as 
$\bar g'{}_{ab} = k^2 \bar g_{ab}$. The quantities $S_{ab}, n^a, f$ change in the following
way:  
\begin{eqnarray}
n^{a \prime} &=& k^{-1} n^a + k^{-2} \Omega \nabla^a k, 
\nonumber\\
S_{ab}' &=& S_{ab} + 2 k^{-1} \nabla_a \nabla_b k 
           - k^{-2}g_{ab}(\nabla_m k) \nabla^m k, \\
f'      &=&  k^{-1} f + 2 k^{-2} n^a \nabla_a k + k^{-3} (\nabla^a k) \nabla_a k
\end{eqnarray}
with similar formulas for the background quantities $\bar S_{ab}, \bar n^a, \bar f$. 
In order to arrive at the above formula for $\Theta$, we assumed that we were in 
a gauge such that $\bar S_{ab} = 0$, and such that $\bar f = const.$
Since we have thereby already partially fixed the gauge, we need to demand 
that $k$ be such that $\bar S'_{ab} = 0$ and such that $\bar f' = const.$
Inserting these formulas, we find that for such $k$
\ben
\label{thetatransf}
\Theta' = \Theta + \delta \Pi, 
\een
where $\Pi$ is the $(d-1)$-form on $\I$ defined by 
\ben
\label{Pidef}
\Pi_{a_1 \dots a_{d-1}} = \frac{(d-2) }{2^6 \pi G}  
\Omega^{-(d-2)} k^{-1} n^b (\nabla_b k)
(g - \bar g)_{cd} q^{ce} q^{df} (g - \bar g)_{ef} 
\, \epsilon_{a_1 \dots a_{d-1}} .  
\een
Since $\delta \Pi$ is not vanishing at $\I$, this means that 
the definition of $\Theta$ is not completely independent upon how 
the physical metric is written as $\Omega^2 \tilde g_{ab}
= g_{ab}$, respectively how the Minkowski metric is written 
as $\Omega^2 \tilde \eta_{ab} = \bar g_{ab}$ in terms of a conformal background metric, 
with $\bar S_{ab} = 0$ and $\bar f = const.$
We resolve this gauge 
ambiguity by chosing a representer $\bar g_{ab}$ in the conformal class of the Einstein static
universe which has 
\ben
\label{gfx}
\bar f = 0, \quad \bar \nabla_a \bar n^b = 0, \quad \bar S_{ab} = 0 \quad \text{near $\I$}, 
\een
where $\bar \nabla_a$ denotes the derivative operator associated with $\bar g_{ab}$.
It can be seen that $\Pi = 0$ for any further gauge change preserving this 
gauge condition, i.e., that $\Theta$ is now defined in a gauge invariant way.  
We will stick with this gauge choice for the remainder of the paper. 
It follows from these gauge conditions that 
\ben
\label{drrr}
f = O(\Omega^{\frac{d}{2}}), \quad \nabla_a n_b = O(\Omega^{\frac{d-2}{2}}). 
\een 
Some further explanations concerning this gauge choice are provided in appendix~B.

By the general analysis reviewed in the last section 
we infer that generators ${\mathcal H}_\xi$ associated 
with asymptotic symmetries $\xi^a$ exist in 
$d$-dimensional general relativity with our choice of asymptotic flatness conditions, 
and that analysis instructs us to define ${\mathcal H}_\xi$ by 
\begin{equation}
\label{neweqn}
\delta {\mathcal H}_\xi = \int_B (\delta Q_\xi - \xi \cdot \theta) + 
\frac{1}{32\pi G} \int_B \tau^{ab} N_{ab} \, \xi \cdot {^{(d-1)} \epsilon}, 
\end{equation}
where $Q_\xi$ and $\theta$ were defined in eqs.~\eqref{nc} and~\eqref{td}.

Formulas~\eqref{neweqn} and~\eqref{Td} are also correct in $d=4$ 
(for a derivation, see~\cite{wz}), provided that $N_{ab}$ is 
given by the usual definition of the news tensor in 4 dimensions, 
$N_{ab} = \zeta^*(q^m{}_a q^n{}_b S_{mn}) - \rho_{ab}$, instead
of eq.~\eqref{newsdef}. Here, $\rho_{ab}$ is the uniquely determined symmetric tensor on $\I$ provided by thm.~5 of~\cite{g}, 
whose precise form depends on the chosen gauge. 

Plugging the expression~\eqref{Td} for the symplectic potential into the flux formula, eq.~\eqref{fluxf}, 
and setting as above $\chi_{ab} = \Omega^{-\frac{d-6}{2}} \pounds_{\xi} \tg_{ab}$
we get the following expression for the flux associated with an asymptotic symmetry $\xi^a$ through a segment $\S$ of $\I$ 
\begin{equation}
\label{fluxd}
F_\xi = \frac{1}{32\pi G} \int_\S \chi^{cd} N_{cd} \, {^{(d-1)} \epsilon},   
\end{equation}
noting that this is finite on account of our definition of an asymptotic symmetry, see eqs.~\eqref{chicond}. 

\medskip

Having established the existence of a generator ${\mathcal H}_\xi$, we now discuss its uniqueness.
The definition of ${\mathcal H}_\xi$ depends on the choice of $\Theta$, which is itself only unique
up to the addition of a $(d-1)$-form on $\I$ of the form $\delta W$, where $W$ is an arbitrary $(d-1)$-form 
on $\I$ that is locally constructed out of the physical metric, the physical Riemann tensor and
its derivatives, and $\Omega$. The change $\Omega \to \lambda \Omega$ and $g_{ab} \to \lambda^2 g_{ab}$ with 
$\lambda$ a constant will keep the physical metric fixed and preserve~\eqref{drrr}, so gauge invariance 
requires that $W \to W$ under this change of the unphysical metric and the conformal factor. (This requirement is 
met by our definition~\eqref{Td} of $\Theta$.) Moreover, the symplectic potential $\Theta$ defined in~\eqref{Td} has the 
property that it vanishes whenever the news tensor, $N_{ab}$, vanishes. A vanishing news tensor
indicates the absence of radiation (at least in 4 dimensions), 
and our definition~\eqref{Td} for $\Theta$ has the property that it vanishes when $N_{ab} = 0$, thereby implying by eq.~\eqref{fluxf}
that the flux also vanishes whenever the news vanishes. 
It is natural to demand that any reasonable definition of $\Theta$,
and hence the flux, vanishes when the news is zero, which in turn
leads to the requirement that $W=0$ whenever $N_{ab} = 0$.  
If $W$ has furthermore an analytic dependence on the (physical) metric, then we claim that 
these requirements imply that $W=0$, and hence that $\Theta$ is unique. 

In order to see that this is indeed true, it is useful to introduce the ``scaling dimension''~\cite{g} of a tensor 
$L^{a\dots b}{}_{c \dots d}$ with $u$ upper indices and $l$ lower indices that is constructed out of the unphysical 
metric and $\Omega$. We say that such a tensor has scaling dimension $s$ if $L^{a\dots b}{}_{c \dots d} \to 
\lambda^{s-u+l} L^{a\dots b}{}_{c \dots d}$ under a change $\Omega \to \lambda \Omega$ and $g_{ab} \to \lambda^2 g_{ab}$. 
It follows from this definition that the scaling dimension does not depend on the position of the indices and is additive
under the tensor product. The dimension
of $g_{ab}$ is $0$, the dimension of $\Omega$ is $+1$, the dimension of the Riemann tensor is $-2$ and each derivative
decreases the dimension by 1, which implies that the dimension of $n_a$ is 0. By assumption, the $(d-1)$ form 
\begin{equation}
W_{ab \dots c} = Y(\Omega, g_{ab}, n_a, \dots, (\nabla_m)^r n_a, R_{abcd}, \dots, (\nabla_m)^t R_{abcd}) 
\, ^{(d-1)} \epsilon_{ab \dots c}
\end{equation}
has scaling dimension $-(d-1)$. Therefore, since $^{(d-1)} \epsilon$ has scaling dimension $0$, $Y$ must have
scaling dimension $-(d-1)$. Using Einstein's equation to eliminate covariant derivatives of $n_a$ in terms of 
covariant derivatives or $S_{ab}$, and using eq.~\eqref{cdefn} to eliminate the Riemann tensor in favor of $C_{abcd}$
and $S_{ab}$, we can write a term in $Y$ schematically in the form
\begin{equation}
\label{gterm}
\Omega^v (n_a)^l
\prod_{i=1}^r (\nabla_m)^{s_i} (\Omega^{-\frac{d-4}{2}} S_{ab}) \prod_{j=1}^u
(\nabla_m)^{t_j} (\Omega^{-\frac{d-6}{2}} C_{abcd}),  
\end{equation}
where we have suppressed contractions with the metric $g_{ab}$ to lighten the notation.
The scaling dimension of this term must be equal $-(d-1)$, which implies that 
\begin{equation}
\label{powercount}
\sum_i s_i + \frac{d}{2}r + \sum_j t_j + \frac{(d-2)}{2}u - v = d-1. 
\end{equation}
Since the expressions $\Omega^{-\frac{d-4}{2}} S_{ab}$ and $\Omega^{-\frac{d-6}{2}} C_{abcd}$ are smooth at $\I$ as a
consequence of our definition of asymptotic flatness, the expression~\eqref{gterm} can be nonvanishing at $\I$ 
if and only if $v \le 0$ (note that $v<0$ is allowed, since the other terms appearing in the above expression 
could vanish at $\I$). Furthermore, $\Omega^{-\frac{d-4}{2}} S_{ab}$ vanishes at $\I$ if and only if the news 
vanishes at $\I$. Therefore, since we want $Y$ to vanish whenever $N_{ab} = 0$, we must have $r > 0$. 
On the other hand, eq.~\eqref{powercount} implies that $r \le 1$, so $r=1$. 
We now analyze the remaining cases: When $r=1$ and $u=0$, 
then the term~\eqref{gterm} looks schematically like
\begin{equation}
\label{lala}
\Omega^v  (n_a)^l (\nabla_m)^{s} (\Omega^{-\frac{d-4}{2}} S_{ab}), 
\end{equation}   
with $s-v = (d-2)/2$. This term has to vanish when the news vanishes and hence when
$\Omega^{-\frac{d-4}{2}} S_{ab}$, but not necessarily its derivatives, vanishes at $\I$. 
This implies $s=0$ and hence $v=-(d-2)/2$, so we need contractions of $n_a$ 
with itself to get a term that is finite at $\I$. But contractions of $n_a$ with itself
give a power of at least $\Omega^{\frac{d+2}{2}}$, therefore terms of the form eq.~\eqref{lala} 
cannot occur. The only remaining nontrivial case is $r=1$ and $u = 1$. 
In this case, we must have $s_i = t_j = v = 0$ and the term~\eqref{gterm} must take the form $\Omega^{-(d-5)} C_{abcd} S^{ac}
n^b n^d$. But this term vanishes at $\I$, by eq.~\eqref{appendid} in the appendix. We have therefore shown that $W = 0$ and 
hence that the symplectic potential $\Theta$ given by eq.~\eqref{Td} is unique under the above assumptions. 
   
\medskip

We now consider the flux for the special case of ``translational'' asymptotic symmetries $\xi^a$. These 
are distinguished by the fact that the restriction of $\xi^a$ to $\I$ is proportional to $n^a$, i.e., 
$\xi^a = \alpha n^a + \Omega k^a$, for some $k^a$, smooth at $\I$. A vector field $\xi^a$ is an asymptotic symmetry
if and only if the tensor $\chi_{ab} = \Omega^{-\frac{d-6}{2}} 
\pounds_{\xi} \tg_{ab}$ satisfies eqs.~\eqref{chicond}. If we substitute this form of $\xi^a$ into eqs.~\eqref{chicond}, 
we see that $k^a = - \nabla^a \alpha$ at $\I$. Let us therefore make the ansatz
\begin{equation}
\label{xiadef}
\xi^a = \alpha n^a - \Omega \nabla^a \alpha. 
\end{equation}
For which $\alpha$ is this an asymptotic symmetry? 
Inserting~\eqref{xiadef} into $\chi_{ab}$, we find that  
\begin{equation}
\label{chipr}
\chi_{ab} = 2\Omega^{-\frac{d-4}{2}}
  \left(
        - \nabla_a \nabla_b \alpha 
        + \Omega^{-1} \alpha \nabla_a n_b 
        + \Omega^{-1} g_{ab} n^c \nabla_c \alpha 
  \right).   
\end{equation}
Using that $\nabla_a n_b = O(\Omega^{\frac{d-2}{2}})$ in our gauge choice (see eq.~\eqref{drrr}) we see that $\chi_{ab}$
is finite at $\I$ if and only if $\alpha$ satisfies 
\begin{equation}
\label{alphadr}
\nabla_a \nabla_b \alpha - \Omega^{-1} g_{ab} n^c \nabla_c \alpha = O(\Omega^{\frac{d-4}{2}}). 
\end{equation} 
Dotting $n^a$ into eq.~\eqref{chipr}, and using that $f = O(\Omega^{\frac{d}{2}})$ in our gauge
choice, we see that $\chi_{ab} n^b = O(\Omega)$ if and only if
\ben\label{alphadr1}
\nabla^b(\Omega^{-1} n^a \nabla_a \alpha) = O(\Omega^{\frac{d-4}{2}}).
\een
Contracting this once more into $n^a$, we see that $\chi_{ab}n^a n^b = O(\Omega^2)$ if 
\ben\label{alphadr2}
\Omega^{-1} n^b \nabla_b (\Omega^{-1} n^a \nabla_a \alpha) = O(\Omega^{\frac{d-4}{2}}). 
\een
In $d>4$, an asymptotic symmetry must furthermore satisfy $\chi^a{}_a = O(\Omega)$. 
However, this condition actually automatically follows for any asymptotic 
symmetry $\xi^a$ once $\chi_{ab} = O(1), \chi_{ab} n^b = O(\Omega), 
\chi_{ab} n^a n^b = O(\Omega^2)$ are satisfied. To see this, we note that since 
$\delta \tg_{ab} = \pounds_{\xi} \tg_{ab}$ 
satisfies the linearized Einstein equation, 
the tensor $\chi_{ab}$ satisfies eq.~\eqref{lee} 
(with $\tau_{ab} = \chi_{ab}$ in that equation). Multiplying 
eq.~\eqref{lee} by $\Omega$, we see that the only remaining singular 
term on the right side is given by a constant that is nonzero for $d \neq 4$, times $\Omega^{-1} n_a n_b \chi_c{}^c$, 
which implies that $\chi_a{}^a = O(\Omega)$ when $d>4$.
Thus, if eqs.~\eqref{alphadr}, \eqref{alphadr1} and \eqref{alphadr2} hold, then 
the vector field $\xi^a = \alpha n^a - \Omega \nabla^a \alpha$ is an asymptotic symmetry.  

The above conditions on $\alpha$ can be understood as follows.
In $d=4$ dimensions, conditions~\eqref{alphadr} and~\eqref{alphadr1}
together imply that $\alpha$ must be constant along the null
generators of $\I$, whereas condition~\eqref{alphadr2} gives 
a restriction on how $\alpha$ is defined off of $\I$. 
Hence, $\alpha$ is essentially an arbitrary function on a given cross 
section of $\I$, which is propagated along the null generator to 
the other cross sections. The corresponding symmetries are commonly 
referred to as ``supertranslations''. They comprise the ordinary 
``pure'' translations, as well as 
an additional infinite set 
of mutually commuting so-called ``angle dependent'' translations.
In $d>4$, the above conditions are more restrictive than in $d=4$ and are analyzed in appendix~C.  
There are now only $d$ linearly independent admissible functions $\alpha$ up to correction
terms which essentially do not affect the restriction of $\chi_{ab}$ to $\I$, in the sense 
that the correction terms do not make a contribution to the flux.
The translational asymptotic symmetries associated with these choices of $\alpha$ 
correspond to the $d$ translational Killing fields in Minkowski spacetime. 
There is no direct analog of the angle dependent translations in higher dimensions.
The asymptotic translations with $\alpha \ge 0$ correspond precisely to the future 
directed\footnote{$\alpha \ge 0$ means that $\xi^a = \alpha n^a$ is
future pointing near $\I^+$.} timelike or null translational Killing 
fields in Minkowski spacetime.

Let us calculate the flux when $\xi^a = \alpha n^a - \Omega \nabla^a \alpha$ is an asymptotic future directed 
time translation, i.e. $\alpha \ge 0$. 
Using Einstein's equation~\eqref{ee} to eliminate the term proportional 
to $\nabla_a n_b$ in eq.~\eqref{chipr} in favor of $S_{ab}$, 
we can bring $\chi_{ab}$ into the form 
\begin{equation}
\chi_{ab} = -\Omega^{-\frac{d-4}{2}}
  \left( 
        2 \nabla_a \nabla_b \alpha 
        + \alpha S_{ab} 
        - 2\Omega^{-1} g_{ab} n^c \nabla_c \alpha 
        - \alpha \Omega^{-1} f g_{ab}
  \right). 
\end{equation}
Substituting it into the flux formula, eq.~\eqref{fluxd}, one finds 
\begin{equation}
\label{fluxf2}
F_{\xi} = -\frac{1}{32\pi G} \int_\S
\alpha N^{cd} N_{cd} \, {^{(d-1)} \epsilon} \le 0. 
\end{equation}
This shows that the flux of energy (defined via {\em any} future directed asymptotic 
time translation) through $\I$ is always negative, 
i.e., that the energy radiated away by the system is always positive. 

\medskip

The generators ${\mathcal H}_\xi$ are determined, in principle, by the defining relation 
eq.~\eqref{neweqn} and the requirement that 
${\mathcal H}_\xi = 0$ on Minkowski spacetime. If $\xi^a$ is not a translation, 
i.e., if $\xi^a$ is tangent to some cross section $B$ of $\I$, then the term involving $\Theta$ vanishes in 
the expression for the variation of ${\mathcal H}_\xi$, eq.~\eqref{neweqn}, and an explicit 
expression for ${\mathcal H}_\xi$
can be derived in basically the same manner as in 4 dimensions, see~\cite{wz}. We will not discuss this case 
here but focus on the case when $\xi^a$ is a translation for the rest of this section. 

In that case, the defining relation~\eqref{neweqn} is not useful to actually find the expression for the
generators $\mathcal H_\xi$, although the right side of that equation is, of course, 
explicitly known. Indeed, in~\cite{wz}, an explicit expression for $\mathcal H_\xi$ in $d=4$ 
was found by verifying that relation~\eqref{neweqn} is satisfied by 
a known expression for the Bondi energy previously given by 
Geroch~\cite{g}. Such a candidate expression is, 
of course, not available in $d>4$, since this is precisely what we are actually looking for
in the first place. 
We therefore proceed by a different route, restricting ourselves for simplicity first 
to the case of the asymptotic translation $\xi^a = \alpha n^a$, with $\alpha$ a constant.  

Consider the $(d-1)$-form $\Theta(\tilde g; \pounds_{\alpha n} \tilde g)$ on $\I$ that is given by 
the integrand of the flux integral, eq.~\eqref{fluxf2}. We extend this to a $(d-1)$-form that is defined
on the entire unphysical spacetime $M$ by setting
\begin{equation}
\Theta_{s_1 \dots s_{d-1}} = 
\frac{1}{32\pi G} \alpha \Omega^{-(d-4)}(S_{ab} S_{cd} q^{ac} q^{bd}) \epsilon_{s_1 \dots s_{d-1}}. 
\end{equation}
Define the vector field 
\begin{eqnarray}
\label{pdef}
 P^a &\equiv& \frac{\alpha}{8(d-3)\pi G} \Omega^{-(d-4)}
      \left( S_{de}  q^{ce}  q^{d[b} n^{a]}\nabla_b l_c 
             - \Omega^{-1}  C^{abcd} n_b l_c n_d
      \right) ,  
\end{eqnarray}
in the interior of the spacetime, where $l_a$ is any smooth vector
field such that 
$l_a l^a = 0$ and $n^a l_a = 1$ on $\I$ and such that relations~\eqref{ldro}
are satisfied. We show in the appendix that 
\begin{equation}
\nabla^a P_a 
= \frac{1}{32\pi G} \alpha \Omega^{-(d-4)} 
S_{ab} S_{cd} q^{ac} q^{bd} + O(\Omega),    
\end{equation}
and it can be verified directly 
from the definition of $P^a$ that $P^a n_a = O(\Omega^2)$. 
Next, define the $(d-2)$-form $\mu$ by\footnote{
Our convention for the $*$ operation of a $p$-form is 
$(* A)_{a_1 \cdots a_{d-p}} = 
\epsilon_{a_1 \cdots a_{d-p}}{}^{b_1 \cdots b_{p}} 
A_{b_1 \cdots b_{p}}$.} 
\begin{equation}
 \mu_{a_1\dots a_{d-2}} 
  = \epsilon_{a_1\dots a_{d-2}cd} l^c P^d 
  = \frac{1}{2}[* (l \wedge P)]_{a_1\dots a_{d-2}}. 
\end{equation}
Then it follows that 
\begin{eqnarray}
\label{dm}
({\rm d}  \mu)_{a_1 \dots a_{d-1}} 
    &=& 2 \nabla_m(P^{[m} l^{n]}) \epsilon_{na_1\dots a_{d-1}} \nonumber\\
    &=& 2n_n \nabla_m (P^{[m} l^{n]}) \epsilon_{a_1\dots a_{d-1}} - 2(d-1)\nabla_m(P^{[m} l^{n]}) n_{[a_1}
\epsilon_{|n|a_2\dots a_{d-1}]} \nonumber\\
&=& (\nabla_m P^m)  \epsilon_{a_1 \dots a_{d-1}} 
  - 2(d-1)\nabla_m(P^{[m} l^{n]}) n_{[a_1}
\epsilon_{|n|a_2\dots a_{d-1}]} + O(\Omega),\nonumber\\ 
&=& (\Theta + {\rm d}\Omega \wedge \varphi)_{a_1\dots a_{d-1}} + O(\Omega),
\end{eqnarray}
where $\epsilon_{a_1 \dots a_{d-1}}$ is as in eq.~\eqref{epsd-1}, where it has 
been used that $P^a n_a = O(\Omega^2)$, and where we have set 
$\varphi_{a_1 \dots a_{d-2}} = 2 \nabla_m(P^{[m} l^{n]})\epsilon_{na_1\dots a_{d-2}}$.

Consider now a segment $\S$ of $\I$ bounded by cross sections $B_1$ and $B_2$, and a sequence of 
smooth $(d-1)$-surfaces $\S_i$ of constant $\Omega$ that approach $\S$. Using 
eqs.~\eqref{fluxf} and~\eqref{dm}, we can write the flux through the segment $\S$ as follows:
\begin{eqnarray}
\label{Fan}
F_{\alpha n} = \lim_{i \to \infty} \int_{\S_i} \Theta(\tilde g, \pounds_{\alpha n} \tg) &=& 
\lim_{i \to \infty} \int_{\S_i} ({\rm d} \mu - {\rm d} \Omega \wedge \varphi) \nonumber\\
&=& \lim_{i \to \infty} \left(
\int_{(\partial \S_i)_2} \mu - \int_{(\partial \S_i)_1} \mu \right) 
\end{eqnarray}
where we have used Stoke's theorem\footnote{A subtlety arises from the fact that the gauge that 
we are working in (chosen such that eq.~\eqref{drrr} holds) is actually not defined on all of $\I$, but only on 
$\I$ minus a single generator, see appendix~B. Therefore, there also ought to appear another ``boundary term''
in eq.~\eqref{Fan} corresponding to that single generator. However, it can be seen 
that this term does not make a contribution by passing to a suitable gauge which is defined on all 
of $\I$, and by transforming the expression for $\mu$ accordingly using formulas very 
similar to those given on p.~50-51 of~\cite{g1}. An example is worked
out in appendix~B. 
Similar remarks apply to other formulas below.} and where we have written $(\partial \S_i)_1$ for the connected
component of the boundary approaching $B_1$ and $(\partial \S_i)_2$ for the connected
component of the boundary approaching $B_2$. Now take the variation of this equation and substitute 
the variation of the flux formula~\eqref{flux}, $\delta F_{\alpha n} = \delta {\mathcal H}_{\alpha n}(B_1)
- \delta {\mathcal H}_{\alpha n}(B_2)$. This gives 
\begin{equation}
\label{diff}
\delta {\mathcal H}_{\alpha n}(B_1)
- \delta {\mathcal H}_{\alpha n}(B_2) 
= \lim_{i \to \infty} \left( 
\delta \int_{(\partial \S_i)_2} \mu 
- \delta \int_{(\partial \S_i)_1} \mu \right). 
\end{equation}
Consider a variation of the metric that vanishes in a neighborhood 
of {\em some} cross section
of $\I$, in addition to satisfying the linearized equations of motion
and the linearized conditions of asymptotic flatness. 
Then it follows from eq.~\eqref{diff} that for such a variation,
\begin{equation}
\label{deltaeq}
\delta {\mathcal H}_{\alpha n}(B) = \delta \int_{B} \mu
\end{equation}
for {\em any} cross section $B$ of $\I$, where the integral on the
right side is defined by the limit of the corresponding integrals 
over $(d-2)$-surfaces of constant $\Omega$ that smoothly approach 
$B$ from the interior of the spacetime\footnote{
This shows in particular that the right side of 
eq.~\eqref{deltaeq} (defined via this limiting procedure) is actually 
finite. This is not obvious from the definition since $P^a$ as well as 
its variation is {\em not} manifestly finite at $\I$ in $d>4$. 
Indeed, the Weyl term in the definition of $P^a$ can only be shown 
to make a contribution of order $O(\Omega^{-(d-4)/2})$ using our
asymptotic flatness conditions. 
}. 
Next, consider a variation that is pure gauge, $\delta \tilde g_{ab}
= \pounds_\eta \tilde g_{ab}$, for some asymptotic symmetry $\eta^a$. 
For such a variation, we have 
\begin{eqnarray}
\delta \int_B \mu &=& \int_B \pounds_\eta \mu \nonumber\\
&=& \int_B [{\rm d}(\eta \cdot \mu) + \eta \cdot {\rm d}\mu]\nonumber\\
&=& \int_B \eta \cdot \Theta(\tilde g; \pounds_{\alpha n} \tilde g), 
\end{eqnarray}
where the integrals are defined by a limiting procedure as above. 
We now show that $\delta {\mathcal H}_{\alpha n}(B)$ for this
variation is also given by the right side of the above equation. For this, 
we consider the 1-parameter family of diffeomorphisms $\Phi_t$ 
generated by $\eta^a$ which maps points in $\I$ to 
points in $\I$. If $\Phi_t^* B$ is the 
cross section of $\I$ obtained from $B$ by applying 
this diffeomorphism, and if $\S_t$ is the segment of $\I$ 
bounded by these two cross sections, then we have 
\begin{eqnarray}
\delta {\mathcal H}_{\alpha n}(B) 
  &=& \lim_{t \to 0} \frac{1}{t} [{\mathcal H}_{\alpha n}(B) - 
{\mathcal H}_{\alpha n}(\Phi_t^* B)]\nonumber\\
&=& \lim_{t \to 0} \frac{1}{t} \int_{\S_t} \Theta(\tilde g; 
\pounds_{\alpha n} \tilde g)\nonumber\\ 
&=& \int_{B} \eta \cdot \Theta(\tilde g; \pounds_{\alpha n} \tilde g), 
\end{eqnarray}
for the variation $\delta \tilde g_{ab} = \pounds_\eta \tilde g_{ab}$, 
where we have used the flux formula eq.~\eqref{fluxf} 
in the second line, and where we have used that 
$\Theta(\tilde g; \pounds_{\alpha n} \tilde g)$ is smooth 
at $\I$ in the third line. 
Hence, we conclude that eq.~\eqref{deltaeq} also holds for 
any variation of the form $\delta \tilde g_{ab} 
= \pounds_\eta \tilde g_{ab}$. Consider now a variation 
with compact support $K$ on some Cauchy surface 
which satisfies the linearized field equations and the linearized 
conditions of asymptotic flatness. 
Then it follows from the arguments given in section~\ref{sect:3} 
that such a variation can be written as a 
sum $\delta \tilde g_{ab} + \pounds_\eta \tilde g_{ab}$, 
where $\eta^a$ is some asymptotic symmetry and 
where $\delta \tilde g_{ab}$ has support in $J^+(K) \cup J^-(K)$. 
It follows that there exist cross sections 
$B^+$ of $\I^+$ and $B^-$ of $\I^-$ such that $\delta \tilde g_{ab}$ 
vanishes in a neighborhood of $B^+$ and 
$B^-$. By the arguments already given, eq.~\eqref{deltaeq} 
therefore holds for any variation which has compact
support on some Cauchy surface. Consequently, eq.~\eqref{deltaeq} 
will also hold for any variation such 
that the corresponding fields $\tau_{ab}, \tau_a$ and their 
first (unphysical) derivatives can be approximated near $B$ 
by the corresponding fields for a sequence of variations 
that have compact support on a Cauchy surface. We believe 
that all variations that satisfy 
the linearized equations of motion and the linearized 
conditions of asymptotic flatness can be approximated 
in this way. Assuming that this is true, it follows that 
eq.~\eqref{deltaeq} holds for all variations. 

We have $\int_B {\bar \mu} = 0$ for our  
background geometry, since we are in a gauge in which $\bar g_{ab}$ is flat
in a neighborhood of $\I$, therefore $\bar \mu = 0$ in 
a neighborhood of $\I$. Therefore, modulo the proof of the
approximation property mentioned in the last paragraph, we have shown that   
\begin{equation}
\label{bm}
{\mathcal H}_{\xi} = \int_B  \mu = \frac{1}{2} \int_B *(l \wedge P)  
\end{equation}
for asymptotic time translations, $\xi^a = \alpha n^a$, 
where $P^a$ given by eq.~\eqref{pdef}. This 
is our expression for the Bondi energy of an asymptotically 
flat spacetime in $d$ dimensions. 

Our expression for the Bondi energy is independent of the particular 
choice of $l_a$ with the properties~\eqref{ldro}: 
Consider another $l_a'$ with the 
same properties as $l_a$, and set $ x_a = l_a - l_a'$. 
Then $ x_a$ satisfies the relations 
$\nabla_b  x_a = O(\Omega^{\frac{d-4}{2}})$, $n_a  x^a =
O(\Omega^{\frac{d}{2}})$ and $l^a  x_a = O(\Omega^{\frac{d-2}{2}})$. 
Consider the antisymmetric tensor field $ X^{ab}$ defined by 
\begin{equation}
X^{ab} = \frac{1}{8(d-3)\pi G} \alpha \Omega^{-(d-4)} 
S_{ef} q^{ed} q^{f[a} n^{b]}  x_d.    
\end{equation}
Then it can be seen, using formulas~\eqref{ldro}, \eqref{drrr} and~\eqref{ger9}, that $ X^{ab} n_b = O(\Omega)$, 
\begin{equation}
\label{pad}
P^{\prime a} =  P^a + \nabla_b  X^{ab} + O(\Omega). 
\end{equation}
Consider now a sequence of embedded $(d-2)$-surfaces that 
are smoothly embedded into 
$M$ and which approach a cross section $B$ of $\I$ as $i \to \infty$. 
Without loss of generality we may assume that $\Omega = const.$ 
on each of these surfaces, so that $n^a$ 
is one of the normals to the surfaces. 
Let $u^a$ be another normal so that the field
$n^{[a} u^{b]}$ is a binormal, meaning that 
$\epsilon_{a_1 \dots a_{d-2}} 
= \zeta^*_i (n^{[m} u^{n]} \epsilon_{mna_1 \dots a_{d-2}})$ 
is equal to the $(d-2)$-volume 
form induced by $g_{ab}$ on each of these surfaces, 
where $\zeta_i^*$ denotes the pull back to $B_i$. 
Then from eq.~\eqref{pad}, we get 
\begin{eqnarray}
\zeta^*_i(\mu -  \mu')_{a_1 \dots a_{d-2}} 
&=& \frac{3}{2}\zeta^*_i \nabla_q (X^{[qm} l^{n]}) \epsilon_{mna_2 \dots a_{d-2}}\nonumber\\
&& -\zeta^*_i\left(\frac{3}{2} X^{[mq} \nabla_q l^{n]} + \frac{1}{2} l^q \nabla_q X^{mn} + P^{\prime [m} x^{n]}
                       \right) n_m u_n \epsilon_{a_1 \dots a_{d-2}}. 
\end{eqnarray}
But the terms in the last line are all of $O(\Omega)$, 
since $X^{ab} = O(\Omega^{\frac{d-4}{2}})$, $X^{ab} n_b = O(\Omega^2)$, $P^a n_a = 
O(\Omega^2)$, and since $x^a n_a = O(\Omega^{\frac{d}{2}})$, 
$n^b \nabla_a l_b = O(\Omega^{\frac{d-2}{2}})$ by eqs.~\eqref{ldro}. 
This shows that, in differential forms notation, 
\begin{equation}
\zeta^*_i (\mu -  \mu') = -\frac{1}{3} \zeta^*_i \{{\rm d}[*(X \wedge l)] \} + O(\Omega),
\end{equation} 
and therefore, by Stoke's theorem, that 
\begin{equation}
\int_B \mu -  \int_B \mu' 
= \lim_{i \to \infty} \int_{B_i} \left( -\frac{1}{3}{\rm d}[*(X \wedge l)] + O(\Omega)\right) = 0, 
\end{equation}
since $B_i$ has no boundary. Thus, our definition~\eqref{bm} of the
Bondi energy does not depend on our choice of $l_a$. 

\medskip

Substituting our expression~\eqref{pdef} for $P^a$ into
eq.~\eqref{bm}, using the definition of $S_{ab}$ together with the
fact that $S_m{}^m = O(\Omega^{\frac{d-2}{2}})$, we can express
${\mathcal H}_\xi$ by the final formula
\begin{multline}
\label{bondimassconst}
{\mathcal H}_\xi = 
\frac{1}{8(d-3)\pi G} \int_B 
\alpha \Omega^{-(d-4)} \Bigg( \frac{1}{(d-2)} R_{ab} 
       q^{ac}  q^{bd} (\nabla_c l_d) n^e l^f \\
     - \Omega^{-1}  l^{[e} C^{f]bcd} n_b l_c n_d \Bigg) 
       \epsilon_{ef a_1 \dots a_{d-2}}. 
\end{multline}
This formula holds for the special translation 
$\xi^a = \alpha n^a$, with $\alpha = const$. 
The above arguments and calculations can be generalized to 
arbitrary (null) translations, $\xi^a = \alpha n^a  
- \Omega \nabla^a \alpha$. One now finds the formula 
\begin{multline}
\label{bondimass}
 {\mathcal H}_\xi = 
 \frac{1}{8(d-3)\pi G} \int_B
 \Omega^{-(d-4)} \Bigg( \frac{1}{(d-2)} R_{ab}  q^{ac}
       q^{bd} (\nabla_c l_d) \xi^e l^f \\ \qquad 
 - \Omega^{-1} \alpha^{-1} (l^{[e} - v \alpha^{-1} \nabla^{[e} \alpha)
 C^{f]bcd} \xi_b (l_c - v \alpha^{-1} \nabla_{c} \alpha) \xi_d \Bigg) 
 \epsilon_{ef a_1 \dots  a_{d-2}},  
\end{multline}
where $v$ is a function such that $\nabla_a v = l_a$.
It can be verified again that this expression does not depend on 
the particular choice of $l_a$. Formula~\eqref{bondimass} can 
alternatively be derived by noting that any null translation 
$\xi^a$ can be obtained from $n^a$ by applying an asymptotic 
symmetry, $\xi^a = \phi^* n^a$. Since the Bondi energy for the vector 
field $\xi^a$ and metric $\tilde g_{ab}$ evaluated at $B$ is 
equal to the Bondi energy for $n^a = \xi^a{}'= \phi^{-1 \,*} \xi^a$ 
and metric $\tilde g'{}_{ab} = \phi^{-1 \, *} \tilde g_{ab}$ 
evaluated at the cross section $B' = \phi(B)$, one can obtain 
the Bondi energy for $\xi^a$ by applying eq.~\eqref{bondimassconst}  
to the metric $\tilde g'{}_{ab}$ and the cross section $B'$. 
The above expression~\eqref{bondimass} is then obtained using 
the formulae $\phi^* g'{}_{ab} = \alpha^{-2} g_{ab}$, 
$\phi^* \Omega = \alpha^{-1} \Omega$, 
$\phi^* l^a{}' = \alpha l^a - v \nabla^a \alpha$, $\phi^* n^a{}' 
= \alpha n^a - \Omega \nabla^a \alpha$, 
as well as our asymptotic flatness conditions. 

Equation~\eqref{bondimass} is the 
main result of our paper. It holds in the gauge defined in~\eqref{gfx}. 
The corresponding formula for other choices of the background geometry 
$\bar g_{ab}$ can be obtained by applying the corresponding gauge 
transformation to our formula. In the case $d=4$, 
formula~\eqref{bondimass} is not correct. An expression in $d=4$ 
has been given by Geroch~\cite{g}. It involves, among other things, 
the news tensor (given by $N_{ab} = S_{ab} - \rho_{ab}$ in $d=4$), 
instead of the unphysical Ricci tensor. 

The first and second term in the integrand of~\eqref{bondimass} 
can be roughly interpreted as follows: the second term is 
the ``Coulomb part'' of the Weyl tensor (multiplied by suitable 
powers of $\Omega$), and represents the ``pure Coulomb contribution'' 
to the Bondi energy. 
The first term represents contributions from gravitational radiation; 
it follows from eq.~\eqref{ldro} given in appendix A that 
it vanishes if and only if the news tensor, $N_{ab}$, and hence the
flux, vanishes. In 4 dimensions, it can be proven~\cite{g} that 
the news tensor, and hence the radiative contribution to the Bondi
energy, always vanishes in stationary spacetimes. It would be 
interesting to see whether an analog of this result holds in $d$
dimensions.

\medskip

In the $d$-dimensional analog of Schwarzschild spacetime given 
by the line element~\eqref{ds}, the Bondi energy is evaluated as
follows. The term involving $R_{ab}$ in our
expression~\eqref{bondimass} for the Bondi energy does not contribute, 
showing that there is no radiative contribution to the Bondi energy. 
The Coulomb contribution is found to be 
$\Omega^{-(d-3)} C^{abcd} l_a n_b l_c n_d = c(d-2)(d-3)/4$ at $\I$. 
Normalizing $\alpha$ so that $\alpha n^a - \Omega \nabla^a \alpha$ 
is equal to the timelike Killing field $t^a$ of the metric~\eqref{ds} 
at infinity gives 
\begin{equation}
\mathcal H_{\alpha n} 
= \frac{c(d-2)A_{d-2}}{16 \pi G} \quad \quad \text{($= \frac{c}{2G}$ 
in 4 dimensions),} 
\end{equation} 
where $A_{d-2}$ is the area of the unit sphere $S^{d-2}$.
This coincides with the ADM mass of the spacetime~\eqref{ds} 
(given e.g. in~\cite{mp}), as we expect.

\section{Conclusions}
\label{sect:6}

We have given a geometrical definition of the asymptotic flatness at null infinity 
in spacetimes of even dimension $d$ greater than $4$ within the framework of conformal infinity. 
Our definition was shown to be stable against perturbations 
to linear order and was shown to be stringent enough to 
allow one to define the total energy of the system viewed from null infinity 
as the generator conjugate to an asymptotic time translation.    
We proposed to take this notion of energy as the 
natural generalization of the Bondi energy to higher dimensions. 
Our definitions of asymptotic flatness and the Bondi energy differ qualitatively 
from the corresponding definitions in $d=4$; although the asymptotic structure of null infinity in higher dimensions 
parallels that in $4$-dimensions in some ways, the latter seems
to be a rather special case on the whole compared to general $d>4$. 

Our definitions and constructions related to asymptotic flatness and Bondi energy do not work in odd spacetime dimensions, 
essentially because the unphysical metric seems to have insufficient regularity properties at null infinity in that case. 
The case of odd dimensional theories of gravity therefore remains open. Apart from this issue, 
the analysis given in this paper could be generalized in two obvious ways: (1) by including matter fields, and (2)
by admitting higher derivative terms such as the square of the scalar curvature in the gravitational action. 

With regard to the first possibility, one would 
first have to formulate appropriate asymptotic conditions on the matter fields, which in practice would presumably 
be found by performing a perturbation analysis. 
We expect the analysis given in section~\ref{sect:3} of this paper to 
generalize  straightforwardly to include conformally invariant fields such as a conformally coupled scalar field, or a 
an abelian $p$-form gauge field $A$ [with Lagrangian density 
$L = {\rm d}A \wedge *({\rm d}A)$] in $d=2p+2$ spacetime dimensions. This 
kind of analysis should also still work for other (non-conformally invariant) massless fields. For massive fields
a different kind of analysis is probably needed, although we expect on physical grounds that these fields 
have the best (i.e., exponential) drop off behavior at null infinity. 
Altogether, we expect that the asymptotic conditions for the 
combined metric and matter fields are given by the conditions given in section~\ref{sect:2} for the metric, plus a condition
of the form $T_{ab} = O(\Omega^s)$ for the stress energy of the matter fields, where $s$ is a suitable number. With 
these conditions in place, a derivation of the Bondi energy can presumably 
be given in close parallel to our analysis in section~\ref{sect:5}. 

With regard to the second possibility, it is much less clear to us what the likely asymptotic conditions on 
the gravitational fields might be in that case, or even how they depend on the actual form of the Lagrangian.
In fact, it is not even clear to us that there will generically be {\em any} reasonable definition of asymptotic flatness
that is stable under linear perturbations. 
Moreover, the linearized equations will now have more derivatives and are therefore presumably harder to analyze than
the linearized Einstein's equations.
One may ignore the issue of stability and simply try to repeat the analysis of this paper and~\cite{wz}
using the asymptotic flatness conditions of section~\ref{sect:2}  
which have been shown to work for general relativity. However, 
even though an expression for the Bondi energy might be found in this way, its physical significance would be far from 
clear under these circumstances.

\vspace{1cm}

{\bf Acknowledgements:} We would like to thank Bob Wald for 
discussions, especially with regard to the 
definition of asymptotic flatness in higher dimensions. 
We also thank Gary Gibbons for conversations. 
We benefited from talks and conversations during 
the YITP workshop, YITP-W-02-19 on ``Extra dimensions and Braneworld'' 
held at the Yukawa Institute for Theoretical Physics at Kyoto University.   
A.I. thanks the Enrico Fermi Institute for hospitality. 
This work was supported in part by NFS grant PHY0090138 to 
the University of Chicago (S.H.) and by the Japan Society for the Promotion 
of Science (A.I.).

\appendix

\section{Derivation of equation~\eqref{pdef}}
\label{sect:appendix}

In this appendix we derive expression~\eqref{pdef} for $P^a$ 
as a solution to the equation 
\begin{equation}
\label{mr}
\nabla^a P_a = 
\frac{1}{32\pi G} \alpha \Omega^{-(d-4)}S_{ab} S_{cd} q^{ac} q^{bd} 
+ O(\Omega). 
\end{equation}

It follows from eqs.~\eqref{1a} and~\eqref{1b}  
that the covector $l_a$ can be chosen in such a way that the following 
conditions are satisfied:
\begin{equation}
\label{ldro}
\nabla_a l_b =  O(\Omega^{\frac{d-4}{2}}), \quad l^a l_a 
             = O(\Omega^{\frac{d-2}{2}}), 
\quad l_a n^a = 1 + O(\Omega^{\frac{d}{2}}), 
\end{equation}
(for example, take $l_a = g_{ab} \bar l^b$, where $\bar l^a$ is a vector field on $\bar M$
such that $\bar g_{ab} \bar l^a \bar l^b = \bar \nabla_a \bar l^b = 0$ and $\bar l^a ({\rm d} \Omega)_a = 1$ in a neighborhood
of $\partial \bar M$). We assume from now on that $l_a$ has been chosen in this way.

>From the defining relation for $S_{ab}$, together with eq.~\eqref{xx}, we have
\begin{eqnarray}
\label{yy}
2n^a \nabla_{[a} \nabla_{b]} l_c 
 &=&  R_{abcd} n^a l^d 
\nonumber\\
 &=&  C_{abcd} n^a l^d + \frac{1}{2}  S_{db} n_c l^d  
     - \frac{1}{2} S_{cb}
     - \frac{1}{2} l_b \nabla_c f  + \frac{1}{2}  g_{bc} l^d \nabla_d f. 
\end{eqnarray}
Contracting this equation with $S_{de}  q^{bd}  q^{ce}$, and making use of the relations
$S_{ab} n^b = -\nabla_a f$, $\nabla_a f = \kappa n_a$, $\kappa = O(\Omega^{(d-2)/2})$ and eq.~\eqref{ldro}, 
gives
\begin{eqnarray}
\label{uu}
\Omega^{-(d-4)} S_{ab} S_{cd} q^{ac} q^{bd} 
 &=& -4\Omega^{-(d-4)} n^a (\nabla_{[a} \nabla_{b]} l_c) 
      S_{de}q^{bd} q^{ce} 
\nonumber \\
 && {} \qquad 
       + 2\Omega^{-(d-4)}  C_{abcf} n^a l^f  S_{de}  q^{bd}  q^{ce} 
       + O(\Omega).   
\end{eqnarray}
Thus, the task is to show that the right side of 
this equation can be written as 
a constant times the divergence of $P^a$, plus terms of order $\Omega$. 

We now evaluate the right side of eq.~\eqref{uu} up to order $\Omega$, 
proceeding term by term and make heavy use of the drop-off 
conditions eq.~\eqref{ldro} and eq.~\eqref{drrr}. 
For the second term in eq.~\eqref{uu} containing the Weyl tensor, 
we have\footnote{This term is of order $\Omega$ 
in 4 dimensions since $C_{abcd}$ itself vanishes at $\I$ 
in 4 dimensions (see thm.~11 of~\cite{g}).}, 
using the symmetry of the Weyl tensor $ C_{abcd} =  C_{cdab}$, 
\begin{eqnarray}
\label{term1}
2\Omega^{-(d-4)}  C_{abcf} n^a l^f  S_{de}  q^{bd}  q^{ce} 
   &=&
      2\Omega^{-(d-4)+1} (\nabla_{[c}  S_{f]b}) l^f  S_{de} q^{bd} q^{ce} 
\nonumber \nonumber \\
   &=& 2\Omega \nabla_{[c} (\Omega^{-(d-4)/2} S_{f]b})l^f  q^{cd}  q^{be} 
       (\Omega^{-(d-4)/2}  S_{de})
\nonumber\\
&& {} 
       + (d-4) \Omega^{-(d-4)} n_{[c}  S_{f]b} l^f  q^{cd}  q^{be}  S_{de} 
\nonumber\\
   &=& - \frac{(d-4)}{2}\Omega^{-(d-4)} 
         S_{ab} S_{cd} q^{ac} q^{bd} + O(\Omega), 
\end{eqnarray}
where the identity [compare eq.~(9) of~\cite{g}] 
\begin{equation}
\label{ger9}
\Omega \nabla_{[a}  S_{b]c} +   C_{abcd} n^d = 0  
\end{equation}
has been used in the first line. We next turn to 
the first term on the right side of eq.~\eqref{uu}. 
This can be written as
\begin{eqnarray}
\label{xxxx}
&& -4\Omega^{-(d-4)} n^a (\nabla_{[a} \nabla_{b]} l_c) 
   S_{de}  q^{bd} q^{ce} 
\nonumber \\
&&{}
   = -4 \nabla_a (\Omega^{-(d-4)}  S_{de} q^{ce} q^{d[b} n^{a]} \nabla_b l_c)
     + 4\Omega^{-(d-4)} (\nabla_{a}  S_{de}) q^{ce}  
       q^{d[b} n^{a]} \nabla_b l_c + O(\Omega) . 
\end{eqnarray}
In the second term on the right side, 
we may replace $\nabla_{a}  S_{ed}$ by 
the expression $2\nabla_{[a}  S_{e]d}$, because 
\begin{eqnarray}
&&\Omega^{-(d-4)} (\nabla_{e} S_{ad}) q^{ce} q^{d[b} n^{a]} \nabla_b l_c 
\nonumber\\
&&{}\,
   = \frac{1}{2}\Omega^{-(d-4)} [(\nabla_e \nabla_d f)  q^{db}  q^{ce} 
     - (\nabla_e  S_m{}^m) q^{ce} n^b
     - 2 \nabla_e(l^a \nabla_a f) q^{ce} n^b] \nabla_b l_c 
     + O(\Omega) \nonumber\\
&&{}\, 
   = O(\Omega), 
\end{eqnarray}
where we have used that 
\begin{equation}
\nabla_a S_m{}^m = \nabla_m S_a{}^m
\end{equation}
by the Bianchi identities, that
$S_m{}^m = O(\Omega^{(d-2)/2})$ and that $\nabla_a f = \kappa n_a$, $\kappa = O(\Omega^{(d-2)/2})$. 
We can now apply the identity~\eqref{ger9} to write the right side 
of eq.~\eqref{xxxx} as  
\begin{eqnarray}
\label{secondt}
&=& -4\nabla_a (\Omega^{-(d-4)}  S_{de}  q^{ce}  q^{d[b} n^{a]} \nabla_b l_c)
    -8\Omega^{-(d-3)} C_{aedf} q^{ce}  q^{d[b} n^{a]} n^f 
     \nabla_b l_c + O(\Omega) 
\nonumber\\
&=&-4\nabla_a (\Omega^{-(d-4)}  S_{de}  q^{ce}  q^{d[b} n^{a]} \nabla_b l_c)
   -4\Omega^{-(d-3)}  C_{aedf} n^a n^f \nabla^d l^e \nonumber\\
&& +4\Omega^{-(d-3)}  C_{aedf} n^a n^f l^e n^c \nabla^d l_c +  O(\Omega) 
\nonumber\\
&=&-4\nabla_a (\Omega^{-(d-4)}  S_{de}  q^{ce}  q^{d[b} n^{a]} \nabla_b l_c)
+4\nabla^d (\Omega^{-(d-3)}  C_{dfea} n^f l^e n^a) \nonumber\\
&& +4\Omega^{-(d-3)}  C_{aedf} n^a n^f l^e n^c \nabla^d l_c
   +4\Omega^{-(d-3)}  C_{aedf} l^e n^f \nabla^d n^a +
O(\Omega),   
\end{eqnarray}
where we have used the tracelessness and symmetries of the Weyl tensor 
and eqs.~\eqref{ldro} and~\eqref{drrr} in the second line, 
and where we have used [compare eq.~(12) of~\cite{g}]  
\begin{equation}
\Omega \nabla^{d}  C_{abcd} +   C_{abcd} n^d = 0  
\end{equation}
in the third line. Using Einstein's equation~\eqref{ee}, 
the last term in the last line of eq.~\eqref{secondt} is seen 
to be equal to $+2\Omega^{-(d-4)} C_{adef} 
n^a l^f  S^{de}$, up to terms of order $\Omega$. 
Using eqs.~\eqref{term1} and eq.~\eqref{ger9}, this term can be further rewritten as
\begin{eqnarray}
  2\Omega^{-(d-4)} C_{adef} n^a l^f  S^{de}
   &=&
       2\Omega^{-(d-4)} C_{adef} n^a l^f S_{bc}q^{bd}q^{ec}
     + 2\Omega^{-(d-4)+1} (\nabla_{[e}S_{f]d})l^fn^el^cq^{bd}S_{bc}
 \nonumber \\
   &&
     - 2\Omega^{-(d-4)+1}(\nabla_{[e}S_{f]d})l^f l^d q^{ec} \nabla_c f
 \nonumber \\
   &=&
     - \frac{(d-4)}{2}\Omega^{-(d-4)} S_{ab} S_{cd} q^{ac} q^{bd}
     + \Omega^{-(d-4)+1} (n^e \nabla_e S_{df})l^fl^c q^{bd} S_{bc}
 \nonumber \\
   &&
     - \Omega^{-(d-4)+1} (\nabla_f S_{de})n^el^fl^c q^{bd} S_{bc}
     + O(\Omega)
 \nonumber \\
   &=&
     - \frac{(d-4)}{2}\Omega^{-(d-4)} S_{ab} S_{cd} q^{ac} q^{bd}
 \nonumber \\
   &&
     + \Omega^{-(d-4)+1} (\nabla_f \nabla_d f + S_{de}\nabla_fn^e)
       l^fl^c q^{bd} S_{bc}
     + O(\Omega)
 \nonumber \\
   &=&
     - \frac{(d-4)}{2}\Omega^{-(d-4)} S_{ab} S_{cd} q^{ac} q^{bd}
     + O(\Omega).
\end{eqnarray} 
The term $+4\Omega^{-(d-3)}  C_{aedf} n^a n^f l^e n^c \nabla^d l_c$ 
on the right side of  eq.~\eqref{secondt} can be seen to be of order $\Omega$
by using the identities
\begin{eqnarray}
   n^c \nabla^d l_c &=& \nabla^d (n^c l_c) - (\nabla^d n^c) l_c =
   O(\Omega^{\frac{d-2}{2}})\\
   C_{aedf}n^a n^f &=&
   - \frac{\Omega}{2}
   (n^a \nabla_a S_{ed} + \nabla_e \nabla_d f + (\nabla_e n^a)S_{ad})
   = O(\Omega^{\frac{d-2}{2}}).
\label{appendid}
\end{eqnarray}
Substituting now eqs.~\eqref{secondt} and~\eqref{term1} 
back into eq.~\eqref{uu}, we obtain 
\begin{equation} 
-(d-3)\Omega^{-(d-4)} S_{ab} S_{cd} q^{ac} q^{bd} 
= 4 \nabla_a \left[\Omega^{-(d-4)}
                     ( S_{de}  q^{ce}  q^{d[b} n^{a]} \nabla_b l_c
                     - \Omega^{-1}  C^{abcd} n_b l_c n_d)  
               \right] + O(\Omega),  
\end{equation}
from which eq.~\eqref{mr} follows immediately.

\section{Conformal gauge choices}     

In this appendix we review transformations related to 
the conformal completion of Minkowski spacetime, thereby eludicating 
our gauge choice~\eqref{gfx} for the background geometry. 
Let us denote by $x^\mu$ the usual Cartesian coordinates of 
Minkowski spacetime $({\mathbb R}^d, \tilde \eta_{ab})$. 
Introducing the radial coordinate 
\ben
r = \sqrt{\sum_{\mu = 1}^{d-1} (x^\mu)^2 }
\een
and $t = x^0$, the Minkowski metric can be rewritten as 
\ben
{\rm d} \tilde s^2 
 = -\d t^2 + \d r^2 + r^2 \d \sigma^2 
 = \Omega^{-2} \{ -\d T^2 + \d \psi^2 + \sin^2 \psi \d\sigma^2 \},   
\een
where $\d \sigma^2$ is the line element of the unit sphere $S^{d-2}$,
and where 
\ben 
\Omega  = 2 \cos\frac{T+ \psi}{2} \cos\frac{T- \psi}{2} . 
\een
The coordinates $T, \psi$ are defined by 
\ben
\frac{T+ \psi}{2} 
 = \tan^{-1} (t+r), \quad  \frac{T- \psi}{2} = \tan^{-1} (t-r).
\een
We view these relations as a map $\lambda$ from 
the portion $\bar M = \{ - \pi < T\pm \psi < \pi, \psi \ge 0\}$ 
of the Einstein static universe ${\mathbb R} \times S^{d-1}$ to Minkowski 
spacetime ${\mathbb R}^d$. In other words, 
\ben
\bar g_{ab} = \Omega^2 \lambda^* \tilde \eta_{ab}, 
\een
where $\bar g_{ab}$ is the metric of the Einstein static universe. 
The boundary of $\bar M$ corresponds to the conformal infinity of 
Minkowski spacetime. It is naturally divided into future/past timelike 
infinity, future/past null infinity $\I^\pm$, 
and spatial infinity. The conformal  factor $\Omega$ is well defined and 
smooth in a neighborhood of the null infinities $\I$, and vanishes there.
The metric $\bar g_{ab}$ is conformally flat (implying $\bar C_{abcd}
= 0$), but not flat, $\bar S_{ab} \neq 0$. 

If we change 
\ben
\bar g_{ab} \to k^2 \bar g_{ab}, \quad \Omega \to k \Omega, 
\een
with $k$ a non-vanishing smooth scalar function defined in a neighborhood 
of $\bar M$ in the Einstein static universe which does not vanish 
at $\I$, then the physical metric remains unchanged, 
and the unphysical metric and the conformal factor remain smooth 
at $\I$.One may use this gauge freedom to make 
suitable ``gauge choices'' for the unphysical metric, and 
we will now discuss some of the choices that are being made 
in the main part of the paper. Let $B$ be a cross section of, say future, 
null infinity which does not intersect spatial infinity. 
Then it is possible to choose $\Omega$ so that $\bar g{}_{ab}$ is 
Minkowskian in an open neighborhood of $B$ not intersecting spatial 
infinity. This can be seen as follows:

Any neighborhood of the indicated form is contained in the causal 
future of some point in the interior of the spacetime, which 
of course, corresponds to the interior of a future directed 
lighcone $V^+$, whose apex we may assume to be at the origin, 
\ben 
  V^+ = \{ x^\mu \;|\; x^\mu x_\mu < 0, x^0 > 0 \}.  
\een  
A conformal factor, $\Omega$, defined on $V^+$ such that 
$\bar g_{ab}$ is flat and Minkowskian and such that the gauge 
condition~\eqref{gfx} is satisfied can be constructed as follows. 
Consider the map $\phi$:  
\bena
\label{phidef}
&&\phi: x^\mu \to x^{\prime \mu} 
  = \frac{a^\mu + b^\mu x^\lambda x_\lambda + 2q^{\mu \nu} x_\nu
         }{2 b^\nu x_\nu}, \nonumber\\
&&a^\mu = (1, -1, 0, \dots, 0), \quad b^\mu = (1, 1, 0, \dots, 0), \quad
q^{\mu \nu} = \eta^{\mu\nu} + a^{(\mu} b^{\nu)},  
\eena
which maps points of the interior of $V^+$ bijectively into points 
in the ``right wedge'' $W$ of Minkowski spacetime 
\ben
  W = \{ x^\mu \;|\; x^1 \geq |x^0| \} .  
\een
The portion of $\I^+$ lying in the causal futuer of $V^+$ 
corresponds, under the map $\phi$, to the ``upper horizon'' of $W$, 
defined by $\partial W^+ =\{ x^\mu \;|\;b^\mu x_\mu = 0,\,x^0 > 0\}$. 
The cross section of $\I^+$ corresponding to 
the lightrays outgoing from the apex of $V_+$ 
corresponds to the ``edge'' ($x^0 = 0 = x^1$) 
of the wedge, whereas the lighrays themselves are represented 
by the null curves generated by $a^\mu (\partial/\partial x^\mu)^a$ 
on the ``lower horizon'' of $W$, given by
$\partial W^-=\{ x^\mu \;|\; a^\mu x_\mu = 0,\,x^0 < 0\}$.  
We find that this map $\phi$ is a conformal isometry 
of Minkowski spacetime with conformal factor 
\ben
\label{omega'}
\Omega = b_\mu x^\mu ,   
\een 
i.e., the background metric 
$\bar g_{ab} = \Omega^2 \phi^* \tilde \eta_{ab}$ is Minkowskian. 
The quantities $\bar f, \bar n^a$ associated with this choice of conformal factor 
are 
\ben
\bar f = 0, \quad \bar n^{a} 
       = b^\mu \left( \frac{\partial}{\partial x^{\mu} } \right)^a,
       \quad 
\bar \nabla_a \bar n^b = 0.
\een  
Thus, the conformal transformation~(\ref{phidef}) with conformal
factor~(\ref{omega'}) satisifies our gauge condition~\eqref{gfx}. 

An awkward feature of the map $\phi$ is that 
it is not globally defined on the boundary of $V^+$, 
for the single null generator corresponding 
to $x^{\prime \mu} = \lambda b^\mu, \lambda >0$ 
of the boundary of $V^+$ is mapped to infinity. Consequently, 
there is also a single corrsponding generator of $\I^+$ which is 
not represented as a corresponding generator of $\partial W^+$, 
or, stated differently, is mapped to the null generator 
at \lq infinity' of the upper horizon $\partial W^+$ 
(corresponding to $b_\mu x^\mu=0,\,x^0>0$ but $x^{\mu = 2,\cdots, d-1} 
\rightarrow \pm \infty$).  
Consequently, the cross sections of $\I^+$ within the causal future of
$V^+$ now corrspond to non-compact cross sections of the upper 
horizon of $W$ (of topolgy ${\mathbb R}^{d-2}$). 
This feature of the conformal embedding $\phi$ has the undesirable 
consequence that the integrals in section~5 over 
cross sections of $\I^+$ inside $V^+$  
are now integrals over a noncompact set and therefore the
convergence issue must be addressed. An example of such an 
integral is $\int_\S \Theta$, where $\Theta$ is the symplectic potential 
$(d-1)$-form introduced above in eq.~\eqref{Td}, and where $\S$ is 
a segment of scri. We will now explain how the convergence issue 
is dealt with in this example.

For this purpose, it is useful to introduce another conformal 
transformation:  
\ben
  \psi: x^\mu \to x^{\prime \mu} 
       = \frac{(x+t)^\nu (x+t)_\nu t^\mu + 2(x + t)^\mu 
             }{(x-t)^\lambda (x-t)_\lambda}, \quad 
 t^\mu = (1, 0, 0, \dots, 0) ,     
\label{def:psi}
\een 
which maps points in $V^+$ into points of the interior of a double cone 
$K$ of Minkowski spacetime, 
\ben
 K = \{ x^{\mu} \;|\; |x^0| + r \leq 1 \} .  
\een 
This map also provides a conformal isometry of Minkowski 
spacetime with conformal factor 
\ben
  \Omega' = -(x - t)^\mu (x - t)_\mu ,    
\label{Omega-psi}
\een
rendering the metric $\bar g'{}_{ab} = \Omega'{}^2 \psi^* \tilde \eta_{ab}$ 
Minkowskian in the portion of spacetime corresponding 
to the interior of the future lightcone $V^+$. 
The derivative operator $\bar \nabla'{}_a$ compatible with 
$\bar g'{}_{ab} = \Omega'{}^2 \psi^* \tilde \eta_{ab}$ is simply equal to 
the coordinate derivative operator $\partial/\partial x^\mu$
associated with Cartesian coordinates. 
The portion of $\I^+$ that can be reached from within $V^+$ 
corresponds precisely to the points $x^\mu$ in $K$ 
such that $\Omega' = 0$, i.e., 
the ``upper cap'' of the double cone $K$, defined by 
$\partial K^+ = \{ x^\mu \;|\; x^0 +r =1,\,x^0>0 \}$. 
Future timelike infinity corresponds to the apex $x^\mu = t^\mu$ of 
the upper cap. 
The apex of $V^+$ corresponds to the apex of the lower cap of $K$ 
given by $x^\mu = -t^\mu$, and the lightrays going out 
from the apex of $V_+$ correspond to the null generators of 
the lower cap, $\partial K^-$, of $K$. 
We should note that this choice of conformal factor 
$\Omega'$ does not satisfy our gauge condition~\eqref{gfx}, 
as the quantities $\bar n^{\prime a}, \bar f'$ associated with $\Omega'$ satisfy 
\ben
\label{nf'}
\bar f' = 4, \quad \bar n^{\prime a} 
 = -2 (x -t)^\mu \left( \frac{\partial}{\partial x^{\mu} } \right)^a,
 \quad 
 \bar \nabla'{}_a \bar n^{\prime b} = -2\delta_a{}^b.
\een
The advantage of the conformal transformation~(\ref{def:psi}) with 
conformal factor $\Omega'$ is however that it 
preserves the compactness of cross sections of $\I^+$. 
In fact, the cross section of $\I^+$ corresponding to the outgoing 
lightrays from the apex of $V^+$ is represented by 
the ``belt'' of $K$, ($x^0 = 0, r=1$), and all other cross sections  
to the future of this particular one are given by cross sections 
of the upper cap, and are therefore topological spheres.  

We compose the two maps $\phi$ and $\psi$ to the map 
$\sigma = \phi \circ \psi^{-1}: K \to W$. Under the map $\sigma$ the
set $\partial W^+$ is identified with the upper cap $\partial K^+$ 
of the double cone $K$. If we denote $\bar g_{ab}$ 
the Minkowskian metric of $W$, then under this map, $\bar g_{ab}$ 
gets mapped to the metric $k^{-2} \bar g'{}_{ab}$, 
where $\bar g'{}_{ab}$ is the 
Minkowskian metric on $K$, and where $k$ is calculated to be 
\ben\label{kdef}
k = (x - t + b)^\mu(x - t + b)_\mu.
\een
The conformal factor $k$ vanishes on the single null generator 
emanating from future timelike infinity 
(represented by the point $x^\mu = t^\mu$ of $\partial K^+$) parallel to
$b^\mu$. This generator corrsponds to the generator at infinity 
in $\partial W^+$ under the map $\sigma$, and the vanishing of $k$ on this 
generator is a reflection of this fact\footnote{ 
In other words, $\partial K^+$ is the comactification of $\partial W^+$.
}. 

Returning to the example integral, 
let $\S$ be a segment of $\I^+$, viewed as a subset of the upper 
horizon $\partial W^+$ of $W$, with non-compact cross-sections, 
i.e., $\S$ has topology ${\mathbb R}^{d-2} \times I$, where $I$ is 
a compact interval. 
Let $\S'$ be the segment of $\partial K^+$ corresponding to $\S$ 
under $\sigma$, i.e., $\sigma(\S') = \S$. 
Then $\S'$ has compact cross sections homeomorphic to $S^{d-2}$ 
as $\partial K^+$ does. 
By eq.~\eqref{thetatransf} the symplectic form satisfies 
\ben 
\label{inttheta}
\int_\S \Theta = \int_{\S'} \Theta' + \delta \int_{\S'} \Pi', 
\een
where $\Theta'$ is given in terms of 
$\Omega' = k^{-1} \sigma^* \Omega$, 
$g'{}_{ab} = k^{2} \sigma^* g_{ab}$ and $\tau'{}_{ab} 
           = k^{\frac{d-2}{2}} \sigma^* \tau_{ab}$ 
by a formula analogous to~\eqref{Td}, 
and where $\Pi'$ is given by eq.~\eqref{Pidef}. Using that 
$n'{}^a = -2(x^\mu - t^\mu) (\partial/\partial x^\mu)^a$ 
by eq.~\eqref{nf'}, as well as the above expression~\eqref{kdef} 
for the conformal factor $k$, one immediately finds that 
\ben
 k^{-1} n'{}^a \nabla'_a k = -2 
\een
at points of $\partial K^+$. Inserting this into the definition of $\Pi'$, 
one gets 
\ben
\Pi' = \frac{(d-2) }{2^5 \pi G}  
\Omega^{\prime -(d-2)} (g' - \bar g')_{cd} q'{}^{ce} q'{}^{df} (g' - \bar g')_{ef} 
\, {}^{(d-1)} \epsilon'.
\een
Now $\Pi'$ and $\Theta'$ are finite at $\I$ as a result of our 
asymptotic flatness conditions, 
and the integrals on the right side of eq.~\eqref{inttheta} are 
over a compact set, $\S'$ of $\I$, 
(viewed as a subset of the upper cap $\partial K^+$ via the map
$\sigma$). This shows that the integral of the symplectic potential
over $\S$ appearing on the left side of eq.~\eqref{inttheta} is convergent. 
The same kind of argument can be made for other integrals 
appearing above in section~5. 

\section{Asymptotic translations} 

We finally discuss the space of translational asymptotic symmetries
$\xi^a$ of the form $\xi^a = \alpha n^a - \Omega \nabla^a \alpha$ in
$d > 4$ dimensions. As discussed in section~5, in order for such a
vector field to be an asymptotic symmetry, we must satisfy 
eqs.~\eqref{alphadr}, \eqref{alphadr1} and~\eqref{alphadr2}, 
which we here repeat for convenience: 
\bena
\label{c1}
\nabla_a \nabla_b \alpha - \Omega^{-1} g_{ab} n^c \nabla_c \alpha 
&=& O(\Omega^{\frac{d-4}{2}}),\\ 
\label{c2}
\nabla^b(\Omega^{-1} n^a \nabla_a \alpha) &=& O(\Omega^{\frac{d-4}{2}}),\\
\label{c3}
\Omega^{-1} n^b \nabla_b (\Omega^{-1} n^a \nabla_a \alpha) 
&=& O(\Omega^{\frac{d-4}{2}}).  
\eena
These equations hold in any conformal gauge choice satisfying
eq.~\eqref{gfx}. Actually, by our asymptotic flatness conditions, 
if these conditions are satisfied for one given asymptotically flat 
metric, they are satsified for any asymptotically flat metric 
(satisfying our gauge choice~\eqref{gfx}). In order to analyze these 
equations, we may therefore choose $g_{ab}$ to be equal to our 
Minkowskian background metric $\bar g_{ab}
= \Omega^2 \phi^* \tilde \eta_{ab}$, where $\phi$ is the conformal 
map $V_+ \to W$ defined above in eq.~\eqref{phidef}, and 
where the conformal factor is defined in eq.~\eqref{omega'}. 
The derivative operator $\bar \nabla_a$ is then given by 
the coordinate derivative operator $\partial/\partial x^\mu$, 
and the associated quantity $\bar n^a$ is given by $b^\mu$
in Cartesian coordinates. Inserting these expressions and going to Cartesian 
coordinates, eqs.~\eqref{c1}, \eqref{c2} and~\eqref{c3} become, 
respectively 
\bena
\label{c4}  
\partial_\mu \partial_\nu \alpha 
  + 2 \eta_{\mu\nu} w^{-1} \partial_v \alpha &=& O(w^{\frac{d-4}{2}})\\ 
\label{c9}
\partial_\mu (w^{-1} \partial_v \alpha) &=& O(w^{\frac{d-4}{2}}) \\
\label{c10} 
w^{-1} \partial_v (w^{-1} \partial_v\alpha) &=& O(w^{\frac{d-4}{2}}), 
\eena
where we have now set $w = x^0 - x^1$ and $v = x^0 + x^1$, and where we 
remember that the location of null infinity corresponds to $\Omega =
-w= 0$ in our gauge. 

Let us first check that the timelike translational Killing vector fields 
\ben
\tau^a = \tau^\mu \left( \frac{\partial}{\partial x^{\mu} } \right)^a, 
\quad \tau^\mu = (\tau^0, \tau^1, \dots, \tau^{d-1}) = const. 
\een
in Minkowski spacetime give rise to solutions of eqs.~\eqref{c4}. 
Under the identification provided by the map $\phi$, these vector 
fields correspond to 
\ben
\xi^a = \phi^* \tau^a = (\alpha b^\mu - \Omega \partial^\mu \alpha ) 
\left( \frac{\partial}{\partial x^{\mu} } \right)^a 
= \alpha \bar n^a - \Omega \bar \nabla^a \alpha, 
\een
where 
\ben
\label{alphadef}
  \alpha = \frac{1}{2}(b^\lambda \tau_\lambda x^\mu x_\mu 
         + 2 q_{\mu\nu} \tau^\mu x^\nu + a^\mu \tau_\mu). 
\een
By construction, since $\phi$ is a conformal isometry of Minkowski 
spacetime, we must have 
\ben
\partial_\mu \partial_\nu \alpha 
 + 2 \eta_{\mu\nu} w^{-1} \partial_v \alpha = 0,   
\een
which can also be verified explicitly. This shows eq.~\eqref{c4}, and the other equations~\eqref{c9}
and~\eqref{c10}  follow by dotting 
$b^\mu$ into this equation. It is not difficult to see that 
(in $d>4$), any other solution $\alpha$ to eqs.~\eqref{c4}, \eqref{c9} and~\eqref{c10} 
is given by 
\ben
\alpha = \alpha_0 + O(\Omega^{\frac{d}{2}}), 
\een
where $\alpha_0$ is given by eq.~\eqref{alphadef}. 
Inserting this into the definition of $\chi_{ab}$, eq.~\eqref{chipr}, 
we see that 
\ben
\label{chiab'}
\chi_{ab} = \chi_{0 \, ab} + O(1) n_a n_b + O(\Omega), 
\een 
where $\chi_{0 \, ab}$ is defined by eq.~\eqref{chipr}, with 
$\alpha$ replaced by $\alpha_0$. Now the integrand of the flux 
associated with an asymptotic symmetry is given by 
(see eq.~\eqref{fluxd}) $N_{ab} \chi^{ab}$ up to numerical 
factors, and $N_{ab} n^b = 0$. 
Thus, the second term in eq.~\eqref{chiab'} does not contribute 
to the flux. This shows that the $\alpha_0$ given 
by eq.~\eqref{alphadef} are essentially the only solutions 
to eqs.~\eqref{c4}, in the sense that any other solution will 
give rise to the same flux. Hence, the vector space of 
infinitesimal asymptotic translations is $d$-dimensional 
in $d>4$, consists of the vector fields 
$\xi^a = \alpha n^a - \Omega \nabla^a \alpha$, with $\alpha$ 
given by eq.~\eqref{alphadef} in the gauge that we are working in. 

Let us finally characterize the $\alpha$ corresponding to a future 
directed timelike translational Killing fields $\tau^\mu$ of 
Minkowski spacetime. 
A point $x^\mu$ of $\I^+$ corresponds to a point on $\partial W^+$ under the map $\phi$, 
so we have $b^\mu x_\mu = 0$ and 
$x^\lambda x_\lambda = q_{\mu\nu} x^\mu x^\nu$ for such points. 
Also, since $\tau^\mu$ is future pointing timelike, 
we have $a^\mu \tau_\mu b^\nu \tau_\nu > 
q_{\mu\nu} \tau^\mu \tau^\nu$. Using this, and the inequality 
obtained by expanding out the relation 
\ben
0\le q_{\mu \nu} [(b^\lambda \tau_\lambda)^{-1/2} \tau^\mu 
  + (b^\lambda \tau_\lambda)^{1/2} x^\mu][
(b^\sigma \tau_\sigma)^{-1/2} \tau^\nu 
+ (b^\sigma \tau_\sigma)^{1/2} x^\nu], 
\een
one easily finds that $\alpha > 0$ on $\partial W^+$.  
Conversely, if $\tau^\mu$ is 
such that $\alpha > 0$ on $\partial W^+$, then one 
sees by the same argument that it must be future directed timelike. 
Thus, future directed timelike translational Killing fields 
correspond to asymptotic symmetries $\xi^a$ 
with $\alpha > 0$ on $\I^+$.

\end{document}